\documentclass[structabstract]{aa}  
\usepackage{graphicx}
\usepackage{txfonts}
\usepackage{natbib}
\begin{document}
   \title{High-mass star formation at high luminosities: W31 at
  $>10^6$\,L$_{\odot}$}


   \author{H.~Beuther
          \inst{1}
          \and
          H.~Linz
          \inst{1}
          \and
          Th.~Henning
           \inst{1}
          \and
          A.~Bik
           \inst{1}
          \and
          F.~Wyrowski
          \inst{2}
          \and
          F.Schuller
          \inst{2}
          \and
          P.Schilke
          \inst{3}
          \and
          S.Thorwirth
          \inst{3}
          \and
          K.-T.~Kim
          \inst{4}
           }
   \institute{$^1$ Max Planck Institute for Astronomy, K\"onigstuhl 17,
              69117 Heidelberg, Germany, \email{beuther@mpia.de}\\
$^2$ Max Planck Institute for Radioastronomy, Auf dem H\"ugel 69,
              53121 Bonn, Germany\\
             $^3$ I. Physikalisches Institut der Univ. zu K\"oln, Z\"ulpicher Str. 77, 50937 K\"oln, Germany\\
             $^4$  Korea Astronomy \& Space Science Institute, 776 Daedeokdaero, Yuseong-gu, Daejeon 305-348, Korea}



\abstract
{High-mass star formation has been a very active field over the last
  decade, however, most studies targeted regions of luminosities
  between $10^4$ and $10^5$\,L$_{\odot}$. In contrast to that, the
  highest mass stars reside in clusters exceeding $10^5$ or even
  $10^6$\,L$_{\odot}$.}
{We want to study the physical conditions associated with the formation
  of the highest mass stars.}
{Therefore, we selected the W31 star-forming complex with a total
  luminosity of $\sim 6\times 10^6$\,L$_{\odot}$ (comprised of at
  least two sub-regions) for a multi-wavelength spectral line and
  continuum study covering wavelengths from the near- and mid-infrared
  via (sub)mm wavelength observations to radio data in the cm regime.}
{While the overall structure of the multi-wavelength continuum data
  resembles each other well, there are several intriguing differences.
  The 24\,$\mu$m emission stemming largely from small dust grains
  follows tightly the spatial structure of the cm emission tracing the
  ionized free-free emission. Hence warm dust resides in regions that
  are spatially associated with the ionized hot gas ($\sim 10^4$\,K)
  of the H{\sc ii} regions.  Furthermore, we find several evolutionary
  stages within the same complexes, ranging from infrared-observable
  clusters, via deeply embedded regions associated with active star
  formation traced by 24\,$\mu$m and cm emission, to at least one
  high-mass gas clump devoid of any such signature. The
  $^{13}$CO(2--1) and C$^{18}$O(2--1) spectral line observations
  reveal a large kinematic breadth in the entire region with a total
  velocity range of approximately 90\,km\,s$^{-1}$. Kinematic and
  turbulent structures are set into context. While the average virial
  mass ratio for W31 is close to unity, the line width analysis
  indicates large-scale evolutionary differences between the southern
  and northern sub-regions (G10.2-0.3 and G10.3-0.1) of the whole W31
  complex. A color-color analysis of the IRAC data also shows that the
  class II sources are broadly distributed throughout the entire
  complex whereas the Class 0/I sources are more tightly associated
  with the active high-mass star-forming regions. The clump mass
  function -- tracing cluster scales and not scales of individual
  stars -- derived from the 875\,$\mu$m continuum data has a slope of
  $1.5\pm0.3$, consistent with previous cloud mass functions.}
{The highest mass and luminous stars form in highly structured and
  complex regions with multiple events of star formation not always
  occurring simultaneously but in a sequential fashion. Warm dust
  and ionized gas can spatially coexist, and high-mass starless cores
  with low-turbulence gas components can reside in the direct
  neighborhood of active star-forming clumps with broad line-widths.}
   \maketitle

\section{Introduction}
\label{intro}

High-Mass star formation research has made significant progress over
the last decade, from a theoretical as well as an observational point
of view.  For recent reviews covering various aspects of high-mass
star formation, we refer to, e.g.,
\citet{beuther2006b,zinnecker2007,bonnell2006,mckee2007,krumholz2008a}.
Unfortunately, because of statistical selection effects and general
low-number statistics at the high-luminosity end of the cluster
distribution, one observational drawback so far was that most (sub)mm
studies of young high-mass star-forming regions targeted sites of
$\leq 10^5$\,L$_{\odot}$, hence stars of mostly less than
30\,M$_{\odot}$ (e.g.,
\citealt{molinari1996,plume1997,sridha,mueller2002,fontani2005}).
However, some notable exceptions exist, e.g., G10.6-0.4
\citep{keto2002a}, G31.41+0.31 \citep{cesaroni1994} or W49
\citep{homeier2005}. Hence, we are lacking solid observational
constraints on the physical and chemical conditions of young high-mass
star-forming regions forming the most luminous and high-mass
stars within our Galaxy. In an effort to overcome these limitations,
we selected one of the most luminous ($\sim6\times 10^6$\,L$_{\odot}$)
but not too distant ($\sim$6\,kpc compared to, e.g., $\sim$11.4\,kpc
for W49) Giant Molecular Cloud/high-mass star formation complex
G10.2/G10.3 (also known as W31) for a detailed multi-wavelength
investigation based on APEX $^{13}$CO and C$^{18}$O mapping
observations, the 875\,$\mu$m ATLASGAL continuum survey
\citep{schuller2009}, Spitzer GLIMPSE/MIPSGAL mid-infrared data
\citep{churchwell2009,carey2009}, and previously published cm
continuum data \citep{kim2002}.

The high-mass star formation complex W31 contains the sub-regions
G10.2-0.3 and G10.3-0.1 which have already been identified as H{\sc
  ii} regions and IRAS sources (IRAS\,18064-2020 and IRAS\,18060-2005)
with large luminosities (e.g., \citealt{woodward1984,ghosh1989}). Its
distance is much debated, and literature values vary between 6\,kpc
\citep{wilson1974,downes1980}, 14.5\,kpc \citep{corbel1997} and
3.4\,kpc \citep{blum2001}. The two main regions are separated by
approximately $15'$ (Fig.~1). At an assumed distance of 6\,kpc, the
total luminosity and gas mass of the region amounts to $\sim6\times
10^6$\,L$_{\odot}$ and $\sim 6\times 10^5$\,M$_{\odot}$
\citep{kim2002}, respectively. The individual luminosities of the main
far-infrared peaks of the two sub-regions G10.2-0.3 and G10.3-0.1 are
estimated to be $\sim 8\times 10^5$\,L$_{\odot}$ and $\sim 6\times
10^5$\,L$_{\odot}$, respectively \citep{ghosh1989}.  Within the
larger-scale H{\sc ii} regions, two ultracompact H{\sc ii} regions
were identified G10.15-0.34 and G10.30-0.15 \citep{wc1989b}. The
luminosities of these two UCH{\sc ii}s based on IRAS data are $\sim
1.5\times 10^6$\,L$_{\odot}$ and $\sim 7\times 10^5$\,L$_{\odot}$,
respectively \citep{wc1989b}, consistent within a factor 2 with the
estimates from \citet{ghosh1989}. It is interesting to note that the
infrared clusters in both regions (G10.2-0.3 and G10.3-0.1) are
spatially offset from the UCH{\sc ii} regions, indicating different
episodes of high-mass star formation.  Furthermore, \citet{walsh1998}
identified four distinct Class {\sc ii} CH$_3$OH maser positions
toward the northern G10.3-0.1 sub-region also known as
IRAS\,18060-2005) but none toward the southern G10.2-0.3 sub-region
(a.k.a.~IRAS\,18064-2020).

While small submm continuum maps of individual clumps within each of
the regions had been obtained over the last few years (e.g., the
SCAMPS project, \citealt{thompson2006}), there did not exist (sub)mm
continuum data encompassing the whole GMC complex.  With the advent of
the 875\,$\mu$m survey of the southern Galactic plane (ATLASGAL,
\citealt{schuller2009}), we now have a complete census of the submm
continuum emission of this large-scale star-forming region.
\citet{kim2002} mapped the whole region in $^{13}$CO(1--0) and
CS(2--1) at a relatively low spatial resolution of $60''$. They reveal
many interesting large-scale features (e.g., that G10.2 and G10.3 may
be located on a shell-like structure). Similarly, \citet{zhang2005c}
observed the complex in C$^{18}$O(1--0) with a poor grid separation of
$1'$. All these studies are not resolving the small-scale structure,
and they are not sensitive to the higher-density gas components.

\section{Observations and archival data} 
\label{obs}

The C$^{18}$O(2--1) and $^{13}$CO(2--1) data at 219.560\,GHz and
220.399\,GHz were observed simultaneously with the Atacama Pathfinder
Experiment (APEX\footnote{This publication is based on data acquired
  with the Atacama Pathfinder Experiment (APEX). APEX is a
  collaboration between the Max-Planck-Institut fur Radioastronomie,
  the European Southern Observatory, and the Onsala Space
  Observatory.}, \citealt{guesten2006}) in July 2009 in the 1\,mm band
over the entire W31 complex in the on-the-fly mode. The APEX1 receiver
of the SHeFI receiver family has receiver temperatures of $\sim$130\,K
at the given frequency \citep{vassilev2008}, and the average system
temperatures during the observations were $\sim$250\,K.  Two
Fast-Fourier-Transform-Spectrometer (FFTS, \citealt{klein2006}) were
connected covering $\sim$2\,GHz bandwidth between 219 and 221\,GHz
with a spectral resolution of $\sim$0.17\,km\,s$^{-1}$. The data were
resampled to 1km\,s$^{-1}$ spectral resolution and converted to main
beam brightness temperatures $T_{\rm{mb}}$ with forward and beam
efficiencies at 220\,GHz of 0.97 and 0.82, respectively
\citep{vassilev2008}.  The average 1$\sigma$ rms of the final spectra
is $\sim$0.98\,K in $T_{\rm{mb}}$. The OFF-positions is approximately
$42'/24'$ offset from the center of the field (selected via the Stony
Brook Galactic Plane CO survey, \citealt{sanders1986}), and it
apparently has emission in a velocity regime between 27.5 and
36.5\,km\,s$^{-1}$ showing up as artificial absorption features in our
data (see section \ref{lines}).  The FWHM of APEX at the given
frequencies is $\sim 27.5''$.

The 875\,$\mu$m data for the region are from the APEX ATLASGAL survey
of the Galactic plane \citep{schuller2009}. The $1\sigma$ rms of the
data is $\sim$70\,mJy\,beam$^{-1}$ and the FWHM $\sim 19.2''$. The
Spitzer IRAC 3.6, 4.5, 5.8 and 8.0\,$\mu$m images as well as the MIPS
24\,$\mu$m are taken from the GLIMPSE and MIPSGAL Galactic plane
surveys respectively \citep{churchwell2009,carey2009}. Furthermore, we
employ the 21\,cm radio continuum observations with a spatial
resolution of $\sim 37''\times 25''$ first published by
\citet{kim2002}.

\section{Results}

\subsection{Multi-wavelength continuum imaging}

\subsubsection{General properties}
\label{general}

\begin{figure*}[htb!]
\includegraphics[angle=-90,width=1.55\textwidth]{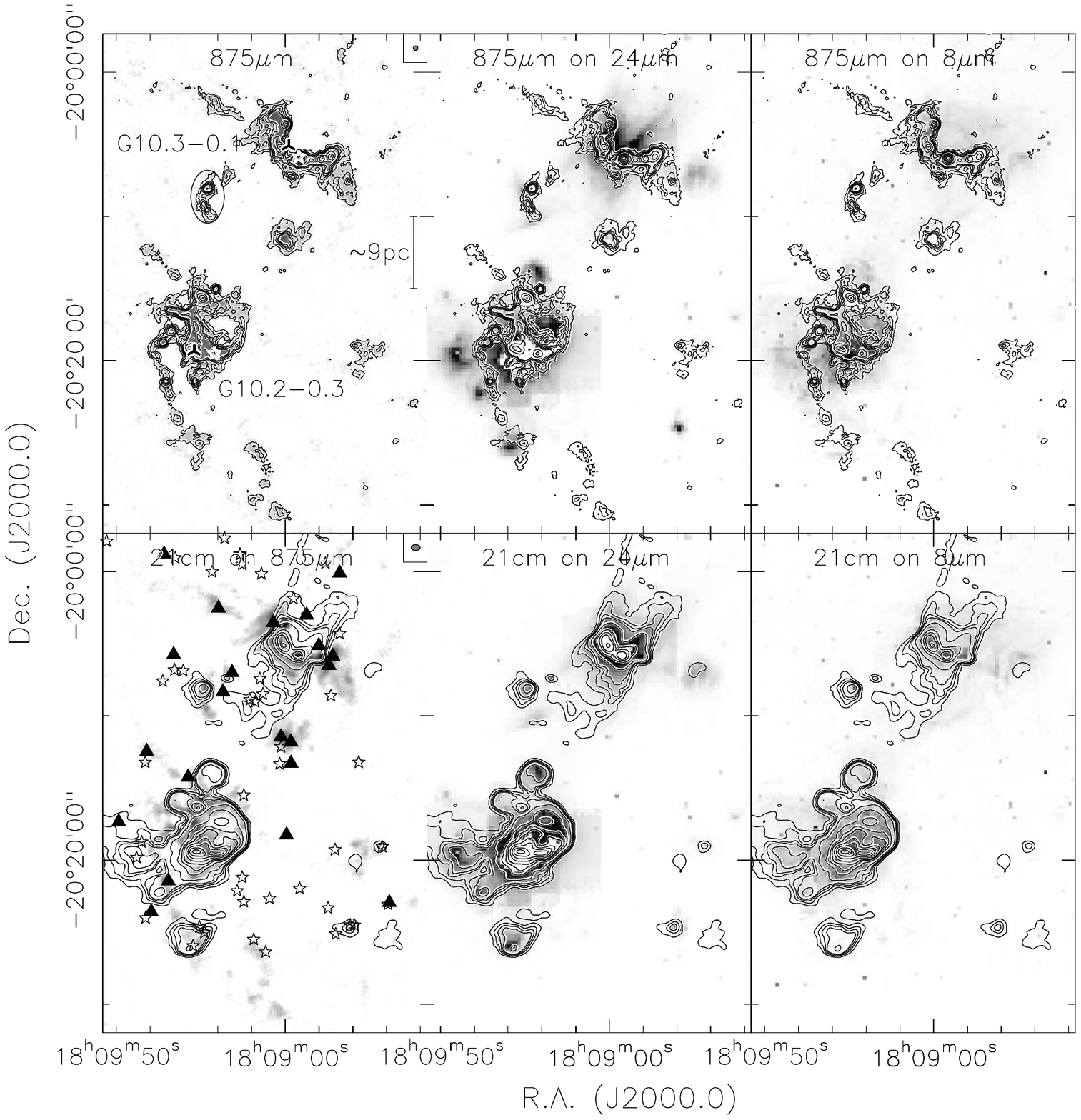}
\caption{Continuum images of the W31 region. The shown wavelengths are
  marked in each panel. The 875\,$\mu$m data are contoured in
  $3\sigma$ steps from 3 to 12$\sigma$ and continue in 15$\sigma$
  steps from 15$\sigma$ onwards ($1\sigma\approx
  70$\,mJy\,beam$^{-1}$). The 21\,cm image is contoured in $3\sigma$
  steps from 3 to 12$\sigma$ ($1\sigma\approx 5$\,mJy\,beam$^{-1}$),
  continuing in 0.24\,Jy\,beam$^{-1}$ steps from 0.12 to
  0.84\,beam$^{-1}$, and then go on in 0.48\,beam$^{-1}$ steps. The
  24\,$\mu$m map is saturated toward the peak positions. In the
  top-left panel the 3-pointed stars mark the position of an O-star by
  \citep{bik2005} and the approximate center of the southern
  infrared-cluster discussed by \citet{blum2001}. The two white
  5-pointed stars show the positions of UCH{\sc ii} regions
  \citep{wc1989b}. The ellipse marks the emission which velocity-wise
  is associated with cloud complexes outside our field of view
  (Sec.~\ref{lines}). A scale-bar is shown in the top-left panel.  The
  875\,$\mu$m and 21\,cm beam sizes are shown in the top-right corners
  of the top-left and bottom-left panels, respectively. In the
  bottom-left panel, the triangles and stars show IRAC-identified
  protostellar class 0/I (the two classes are combined) and class II
  candidates, respectively, following
  \citet{allen2004,megeath2004,qiu2008}.}
\label{continuum}
\end{figure*}

Figure 1 shows a compilation of the different continuum datasets
analyzed in this work ranging from the mid-infrared with IRAC
8\,$\mu$m and 24\,$\mu$m data, to the submm regime at 875\,$\mu$m and
then to radio wavelengths at 21\,cm. In a simplified picture, we
selected the mid-infrared images to trace warm dust emission, the
submm data to study cold dust emission, and the cm observations to
investigate the ionized gas components. We clearly identify the
large-scale structure of the two star formation and H{\sc ii} region
complexes G10.2-0.3 in the south and G10.3-0.1 in the north. At all
wavelengths the two complexes exhibit significant sub-structure,
probably most strongly recognizable in the 875\,$\mu$m image. Since we
are dealing with two very luminous but already relatively evolved star
formation complexes it is not surprising that many emission
features are visible at all presented wavelengths. They show ionized
gas as well as warm dust emission caused by the luminous O and B stars
\citep{blum2001,bik2005}, and furthermore a large amount of cold dust
emission from the still existing gas/dust reservoir present in the
whole region. While the former signifies that star formation is
already ongoing for a considerable amount of time ($>10^5-10^6$\,yr),
the latter shows that the original gas cloud is not yet dispersed and
may still allow further accretion.

\begin{figure}[htb]
\includegraphics[angle=-90,width=100mm]{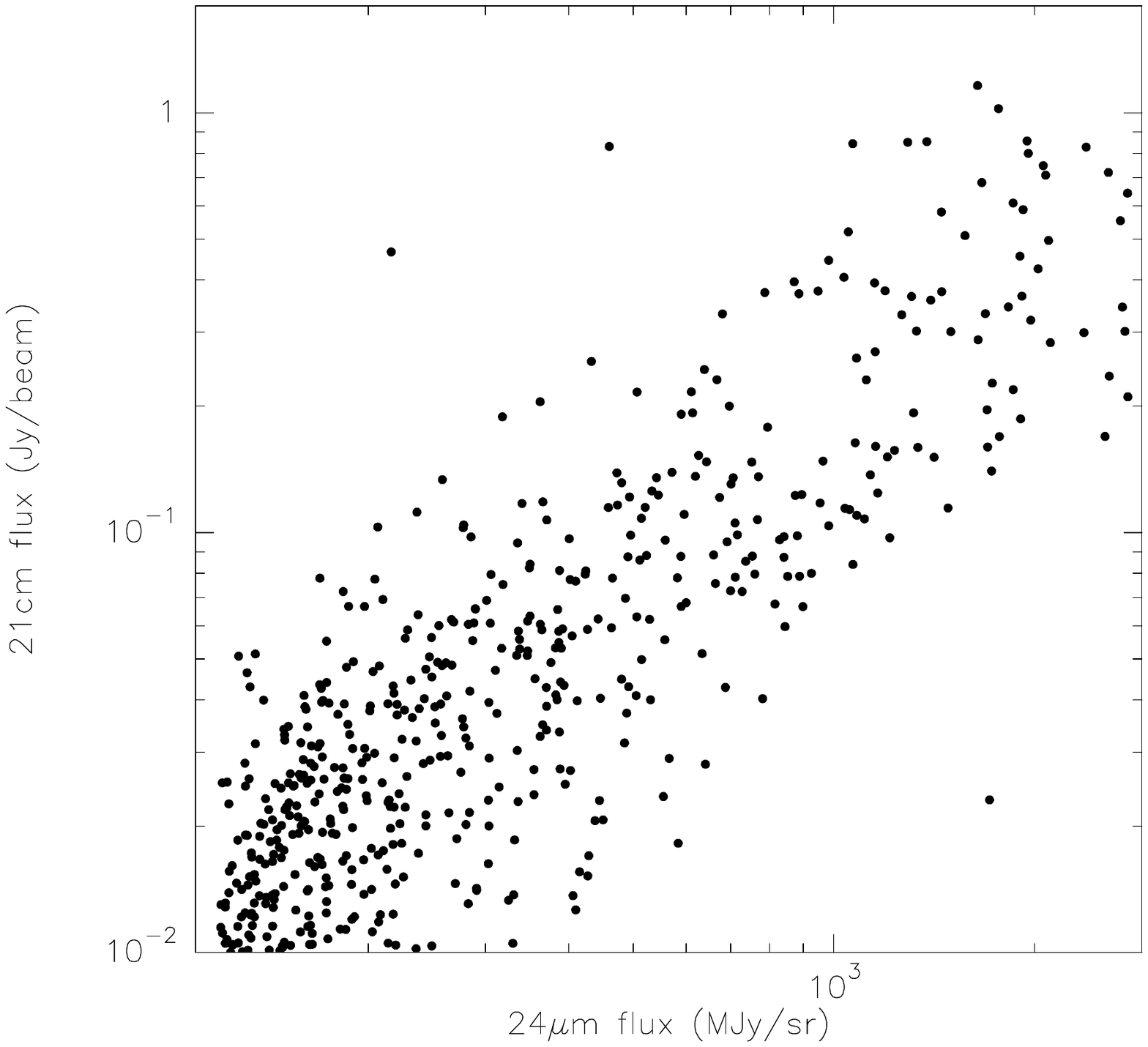}
\caption{Correlation between 24\,$\mu$m fluxes and 21\,cm fluxes for
  the W31 complex. Only pixels with 24\,$\mu$m flux above
  120\,MJy\,sr$^{-1}$ (corresponding to the approximate outskirts of
  the H{\sc ii} regions) and 21\,cm emission above the $3\sigma$ level
  of $\sim$0.01Jy\,beam$^{-1}$ are plotted. Furthermore, only
  correlations on scales of the beam size of the 21\,cm observation
  were used (2-pixel steps corresponding to $29''$).}
\label{scatter}
\end{figure}

Furthermore, the 24\,$\mu$m and 21\,cm data spatially resemble each
other well, for example the cometary shape of the northern region
G10.3-0.1. An even clearer representation of the correlation between
21\,cm and 24\,$\mu$m emission can be achieved if one plots the
observed fluxes pixel by pixel. Figure \ref{scatter} presents such a
``scatter plot'' correlating all pixels with 24\,$\mu$m flux above
120\,MJy\,sr$^{-1}$ (corresponding to the approximate outskirts of the
H{\sc ii} regions) and the 21\,cm emission above the $3\sigma$ level
of $\sim$0.01Jy\,beam$^{-1}$. Only scales of the order the beam size
of the 21\,cm observations were correlated (2-pixel steps
corresponding to $29''$).  This figure outlines the good spatial
correlation between warm dust and ionized gas emission over a broad
range of fluxes in the W31 complex. Hence warm dust below the
sublimation temperature of $\sim$1500\,K spatially coexists with much
warmer ionized gas of temperatures around $10^4$\,K (e.g.,
\citealt{helfand2006}). Recently, \citet{everett2010} argue that such
dust may stem from continuous ablation processes of small cloudlets
that have not been entirely destroyed by the H{\sc ii} region yet.
However, the spatial resolution of the 21\,cm observations corresponds
to linear scales of about $\sim 0.9$\,pc.  At these scales it is also
possible that the observed 21\,cm/24\,$\mu$m correlation stems from an
interface between the edges of H{\sc ii} regions and dense dust shells
which are just not resolved by the observations. One should keep in
mind that the 24\,$\mu$m band is mainly sensitive to very small,
stochastically heated dust grains (e.g.,
\citealt{draine2003,carey2009}) and that, therefore, strong 24\,$\mu$m
emission does not necessarily imply large dust column densities.

\begin{figure*}[htb]
\includegraphics[angle=-90,width=184mm]{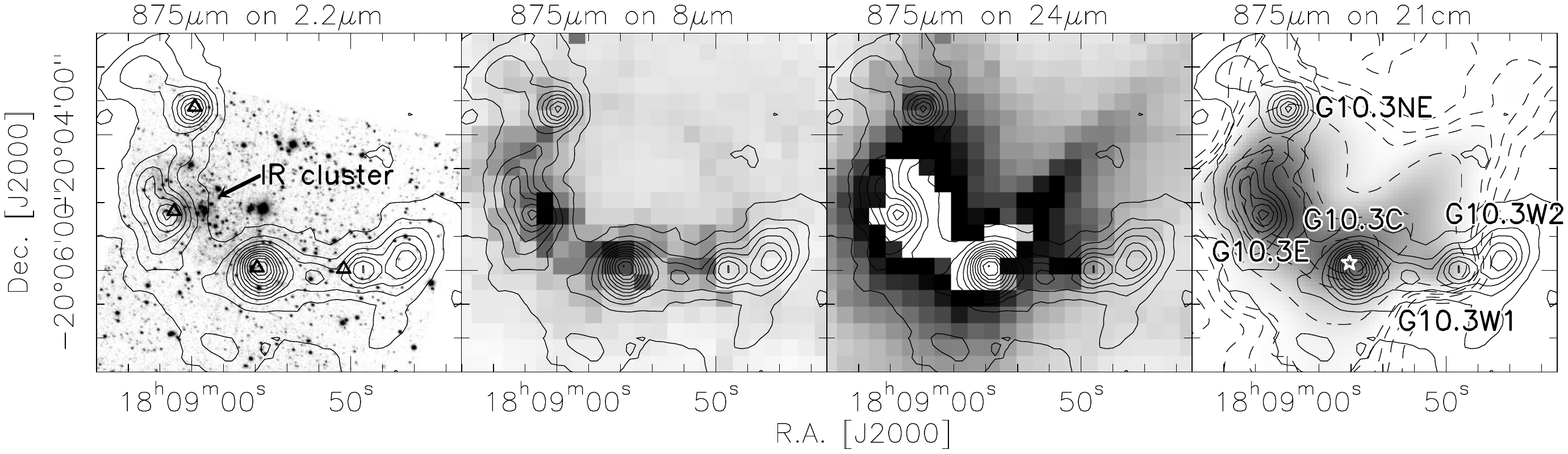}
\caption{Zoom into the G10.3 complex. In all 4 panels the solid
  contours present the 875\,$\mu$m continuum contours (start at
  $3\sigma$ and continue in $9\sigma$ steps, $1\sigma\approx
  70$\,mJy\,beam$^{-1}$) whereas the grey-scale shows other wavelength
  data as outlined over each panel. The dashed 21 cm contours in the
  right panel go from 15 to 60\,mJy\,beam$^{-1}$ in
  15\,mJy\,beam$^{-1}$ steps, and from 120 to 840\,mJy\,beam$^{-1}$ in
  240\,mJy\,beam$^{-1}$ steps. The K-band data are taken from
  \citet{bikphd}. The white central part of the 24\,$\mu$m image is
  saturated. The triangles in the left panel mark the Class {\sc ii}
  CH$_3$OH maser positions from \citet{walsh1998}.  The white star in
  the right panel marks the position of an UCH{\sc ii} region
  \citep{wc1989b}, and the additional labels there name sources
  discussed in the main text.}
\label{zoom}
\end{figure*}

While in many cases we see clear associations of emission at different
wavelengths, we also witness exactly the contrary. Figure \ref{zoom}
presents a zoom into the northern complex G10.3-0.1, where we also
added the near-infrared K-band data from \citet{bikphd}.  The latter
outline the location of the infrared cluster that is offset from the
emission we see at other wavelengths. The O-star marked in
Fig.~\ref{continuum} \citep{bik2005} is part of that cluster. Directly
east of the infrared cluster we find the associated submm clump G10.3E
(clump 4 in Table \ref{clump_parameters_smooth}) and further to the
north G10.3NE (clump 2 in Table \ref{clump_parameters_smooth}), both
associated with mid-infrared, cm and Class {\sc ii} CH$_3$OH maser
emission indicating active star formation. Moving west from the
infrared cluster we identify a clear emission peak at all wavelengths
(G10.3C in Fig.~\ref{zoom} and clump 1 in Table
\ref{clump_parameters_smooth}) marking already a relatively evolved
part of the UCH{\sc ii} region.  Going west, the next source G10.3W1
is still associated with mid-infrared, cm continuum and Class {\sc ii}
CH$_3$OH maser emission (G10.3W1 and G10.3W2 merge into clump 6 in the
smoothed data of Table \ref{clump_parameters_smooth}).  Although at
a weaker level, this also indicates active high-mass star formation
from this gas clump.  However, in strong contrast to these
sub-regions, the most western submm peak G10.3W2 in Figure \ref{zoom}
shows comparably strong submm continuum emission but a complete
absence of any 24\,$\mu$m and 21\,cm emission.  Hence this is a
high-mass gas clump at a very early evolutionary stage, potentially
still in a starless phase prior to any star formation activity. The
projected separation of the G10.3W1 and G10.3W2 peak position of $\sim
42''$ corresponds at the assumed distance of 6\,kpc still to a
projected spatial linear separation of $\sim$1.2\,pc. Hence each of
the submm emission peaks covers scales corresponding to clusters, and
therefore each may form a smaller cluster~-- likely also containing
high-mass stars~-- within the large-scale environment of W31.

Figure \ref{continuum} also shows the distribution of protostellar
class 0/I (the two classes are combined) and pre-main-sequence class
II sources identified by Spitzer IRAC data based on the selection
criteria developed by \citet{allen2004}, \citet{megeath2004} and
\citet{qiu2008}.  While the non-detection of any source toward the
centers of the two sub-regions is likely an artifact due to saturation
and confusion in these areas, one can tentatively identify a trend
that the class 0/I sources are more closely associated with the
centers of activity than the class II sources which appear more widely
distributed over the field of view. Quantitatively speaking, the mean
separations of the northern class 0/I and class II sources from the
UCH{\sc ii} region G10.30-0.15 are $\sim 321''$ and $\sim 428''$,
respectively, whereas the corresponding mean separations for the
southern sources with respect to the UCH{\sc ii} region G10.15-0.34
are $\sim 936''$ and $\sim 1048''$, respectively. These data also
indicate that the southern cluster is spatially more distributed and
may potentially be already more evolved (see section \ref{evolution}).

\subsubsection{Clump properties of the 875\,$\mu$m continuum emission}
\label{clump_prop}

We can now locate the dense 875\,$\mu$m gas and dust clumps at a
spatial resolution of $19''$, corresponding to linear scales of
$\sim$0.5\,pc at an assumed distance of 6\,kpc. While two of the dense
cores are associated with the ultracompact H{\sc ii} regions
G10.15-0.34 and G10.30-0.15 (Fig.~\ref{continuum}), there are
obviously many dense gas clumps that are promising candidates of
ongoing and potentially even younger high-mass star formation. At the
given spatial scales of our resolution limit, each submm sub-source
does not form individual stars but they all should be capable to form
sub-clusters within the larger-scale region.

\begin{figure*}[htb]
\includegraphics[angle=-90,width=1.3\textwidth]{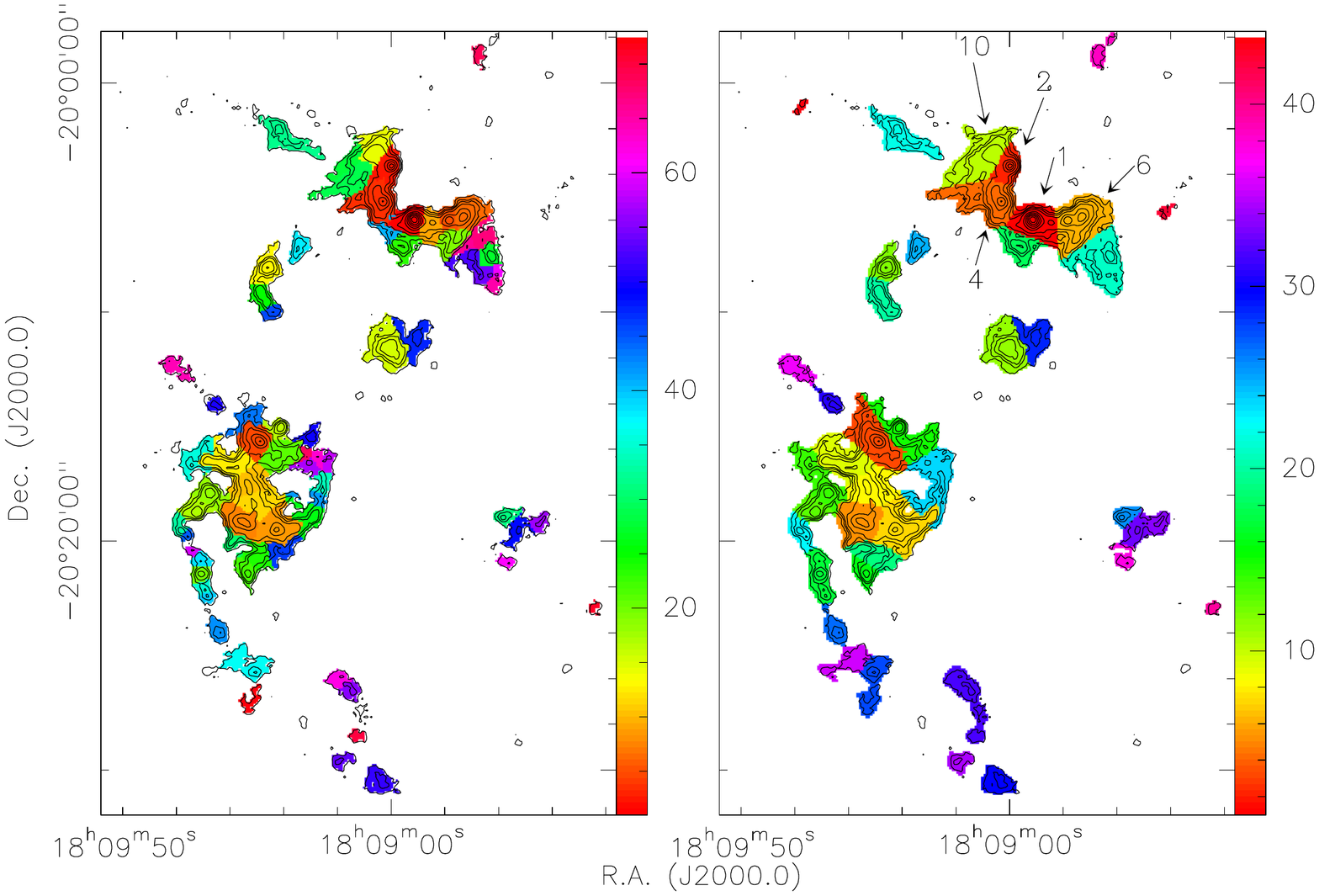}
\caption{The color-scales show the clump structure and boundaries
  derived with the automized clumpfind procedure on the original
  (left) and smoothed (right) 875\,$\mu$m continuum observations. The
  contours show the original 875\,$\mu$m observations with the same
  contour levels as in Fig.~\ref{continuum}. The wedge reflects the
  clump numbers as in Tables \ref{clump_parameters} \&
  \ref{clump_parameters_smooth}. A few discussed clumps are labeled in
  the right panel (see also Fig.~\ref{masses}).}
\label{clumps}
\end{figure*}

To systematically identify gas and dust clumps and to study their
properties in such a large and complex region we employ the clumpfind
algorithm \citep{williams1994} with $3\sigma$ contour levels
(210\,mJy\,beam$^{-1}$). With this procedure we identify 73 submm
continuum clumps over the entire W31 region (Fig.~\ref{clumps}). These
clumps are listed in Table \ref{clump_parameters} with the spatial
coordinates of their peak positions, their integrated and peak fluxes,
and their linear effective radii $r_{\rm{eff}}$.  Assuming that the
submm emission stems from optically thin dust, we can calculate the
H$_2$ gas column densities and masses following
\citet{hildebrand1983,beuther2002a,beuther2002erratum}.  Since we do
not know the temperature sub-structure of the entire complex, we use
an average dust temperature of 50\,K that should be a reasonable proxy
for such an active region (e.g., \citealt{sridha}).  Furthermore, we
use a dust grain emissivity index of $\beta = 2$ (corresponding to a
dust absorption coefficient of $\kappa_{875}\approx
0.8$\,cm$^2$\,g$^{-1}$), and a gas-to-dust mass ratio of 186 is
assumed \citep{jenkins2004,draine2007}.  Table \ref{clump_parameters}
lists also the derived H$_2$ column densities and masses of the
respective gas clumps. With the given uncertainties of temperature and
dust composition, we estimate the masses and column densities to be
accurate within a factor $\sim$3.

The derived clump masses range between 120 and 8200\,M$_{\odot}$ with
a total gas mass in the region of approximately $1.2\times
10^5$\,M$_{\odot}$.  The peak column densities vary between $1.5\times
10^{22}$ and $3.4\times 10^{23}$\,cm$^{-2}$. While the proposed
threshold for high-mass star formation of 1\,g\,cm$^{-2}$
\citep{krumholz2008b} corresponds to column densities of $\sim 3\times
10^{23}$\,cm$^{-2}$, this does not imply that most clumps are not
capable of high-mass star formation, the data clearly show the
opposite. In contrast to this, the calculated column densities are
measured with a linear beam size of $\sim$0.5\,pc, and hence only the
average column densities over such area have the derived values. The
intrinsic column densities at smaller spatial scales are significantly
higher assuming a typical density distribution $\propto r^{-2}$ (see
also \citealt{vasyunina2009} for similar estimates conducted for
infrared dark clouds).

This implies that a large number of the gas clumps in this region
should be capable of forming sub-clusters containing high-mass stars.
To estimate how much mass a gas clump needs to have to form a high-mass
star, we produce stellar cluster toy-models assuming a star formation
efficiency of 30\% and an initial mass function (IMF) following
\citet{kroupa2001}. To form at least one high-mass star of, e.g., 20
or 40\,$M_{\odot}$, in this scenario the initial gas clumps have to
have masses of $\sim 850$ or $\sim 1900$\,M$_{\odot}$, respectively.
Similar results were recently obtained observationally by
\citet{johnston2009}. This clearly indicates that a significant
fraction of gas clumps in the W31 region is capable to form stars in
excess of 20\,$M_{\odot}$.

\begin{figure}[htb!]
\includegraphics[width=90mm]{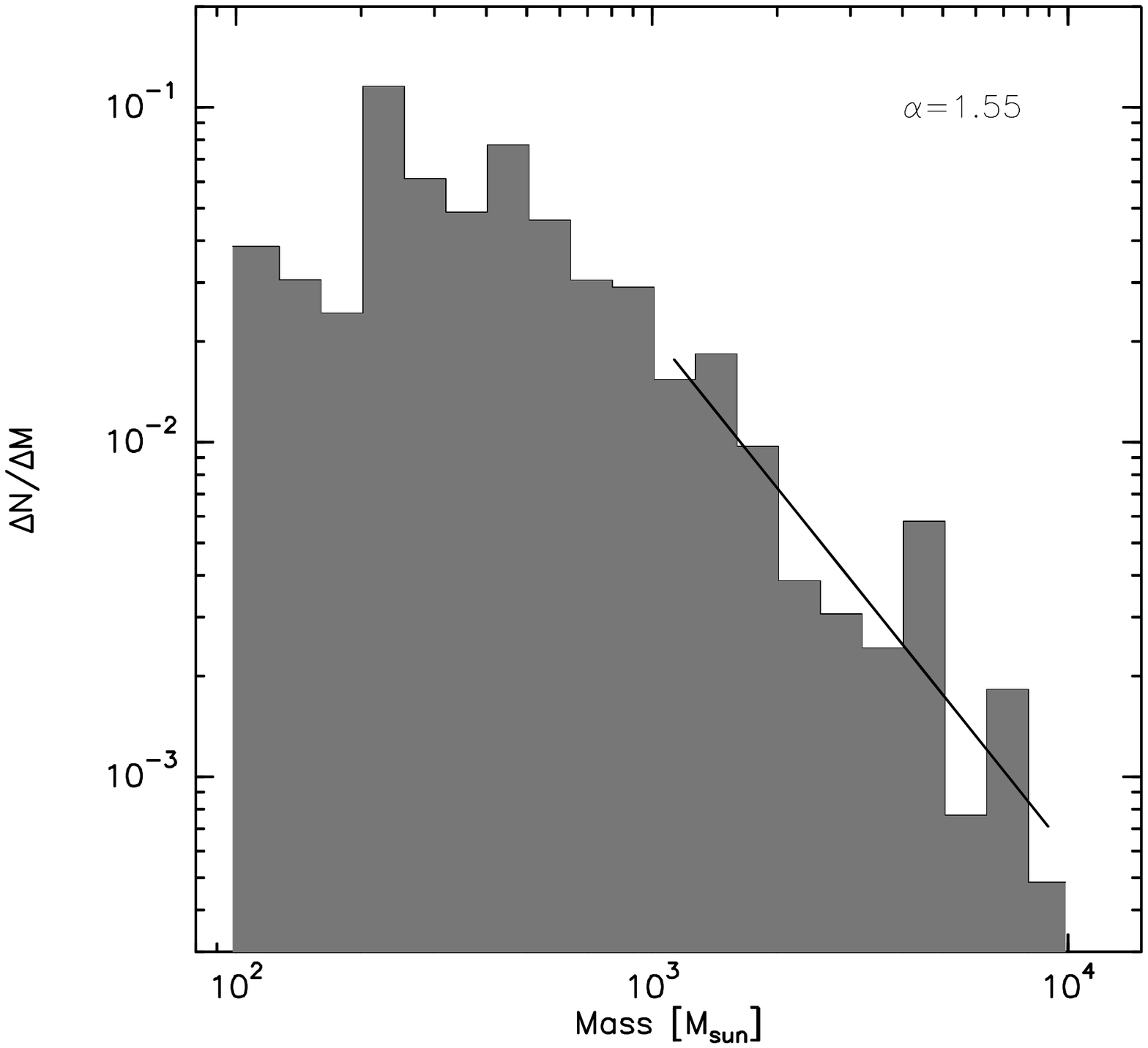}
\caption{Example clump mass function for the W31 complex for one bin
  size (here logarithmic bins with $10^{k*0.1}<M<10^{(k+1)*0.1}$ with
  $k$ an integer, 36 different bin sizes have been used to overcome
  fitting artifacts).  Only clumps with masses $>1000$\,M$_{\odot}$
  were used for the fit.}
\label{cmf}
\end{figure}

Using the clump masses from Table \ref{clump_parameters} we also tried
to derive a clump mass function $\Delta N/\Delta M \propto
M^{-\alpha}$ for the region. To overcome fitting artifacts due to
different bin sizes, we fitted the data, systematically changing the
bin width (see also Rodon et al.~subm.~to A\&A). To avoid any
sensitivity cutoff we fitted only clumps above 1000\,M$_{\odot}$ and
furthermore required at least three non-empty bins.  This procedure
resulted in 36 fits to the data with different bin sizes
each time, allowing us to asses better the error margins of this
approach. Figure \ref{cmf} shows one example fit. While the Poisson
errors for each individual fit are relatively large, the combined
assessment of all different fits from varying bin sizes in this
fitting procedure as well as a comparison to a cumulative fit, allows
us to derive a power-law distribution index within reasonable error
margins. Our derived power-law index $\alpha$ from this procedure is
$\sim 1.5\pm 0.3$ (see discussion in section \ref{cmf_discussion}).

\subsubsection{Smoothed 875\,$\mu$m data}
\label{smooth}

For a better comparison with the spectral line data discussed in the
following section, we also smoothed the 875\,$\mu$m continuum data to
the same spatial resolution as the spectral line observations ($\sim
27.5''$). On this smoothed map with a lower rms of
50\,mJy\,beam$^{-1}$, we again apply the clumpfind algorithm to
extract sources, resulting in 44 clumps compared to the 73 clumps in
the original data (see Fig.~\ref{clumps} for comparison). For these
clumps we also calculated masses and column densities as conducted in
the previous section for the un-smoothed data.  In addition for
comparison purposes with the spectral line data, we also calculate the
mass just within the central beams toward each peak position
$M_{\rm{peak}}$. All the parameters extracted for the smoothed clumps
are listed in Table \ref{clump_parameters_smooth}.
\subsection{Kinematics from spectral line data}
\label{lines}

Star formation processes significantly shape the kinematic and dynamic
properties of the molecular gas. While early-on dense starless gas and
dust clumps usually exhibit narrow line widths because no
(proto)stellar feedback has yet altered the original gas
properties, during ongoing star formation, the natal gas
clumps are strongly influenced by the central star-forming processes
which is reflected in the observable spectral signatures.

\begin{figure}[htb]
\includegraphics[angle=-90,width=90mm]{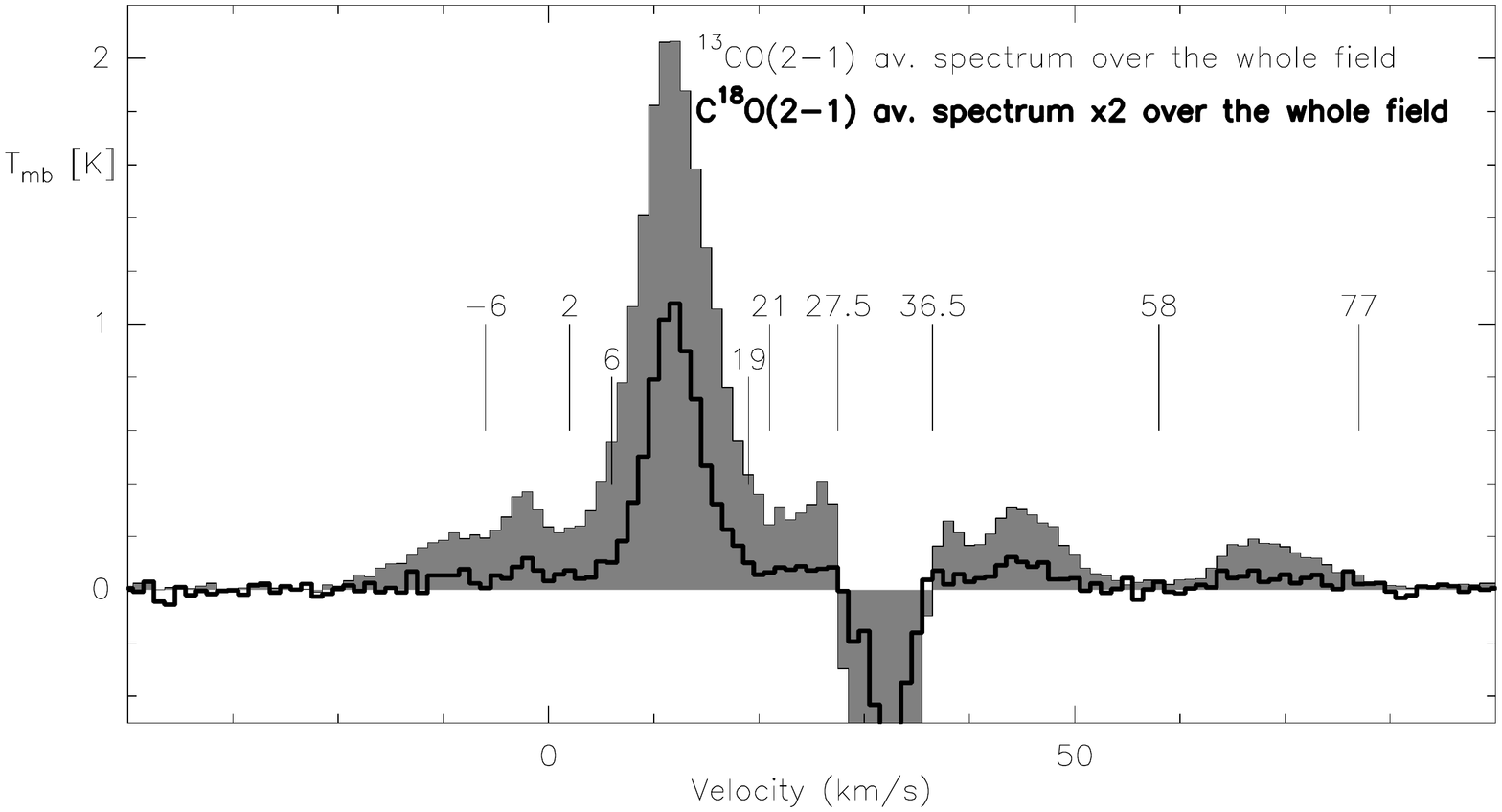}
\caption{$^{13}$CO(2--1) and C$^{18}$O(2--1) (multiplied by a factor
  2) spectra averaged over the entire W31 region. The marked
  velocities show the different velocity regimes used for the
  $^{13}$CO (upper) and C$^{18}$O (lower) moment maps.}
\label{spectra}
\end{figure}

The $^{13}$CO(2--1) and C$^{18}$O(2--1) observations allow a kinematic
analysis of the region. Figure \ref{spectra} presents the spectra of
both species averaged in each case over the entire field of view of
the observations. In particular the $^{13}$CO(2--1) data show the very
broad velocity range present in the region. The figure also marks
different velocity regimes used by us to produce moment maps of the
data (see below). The velocity regime from 27.5 to 36.5\,km\,s$^{-1}$
appears in absorption in this averaged spectrum. This, however, is an
artifact because we see it in all spectra and hence it stems from
emission at this velocity in the OFF position. Interestingly, this
velocity range is exactly the same as the velocities we find emission
for toward the submm continuum peak marked by an ellipse in Figure
\ref{continuum}. Hence, this continuum feature in our maps should be
associated with other cloud complexes outside of our observed field of
view.

\begin{figure*}[htb]
\includegraphics[angle=-90,width=190mm]{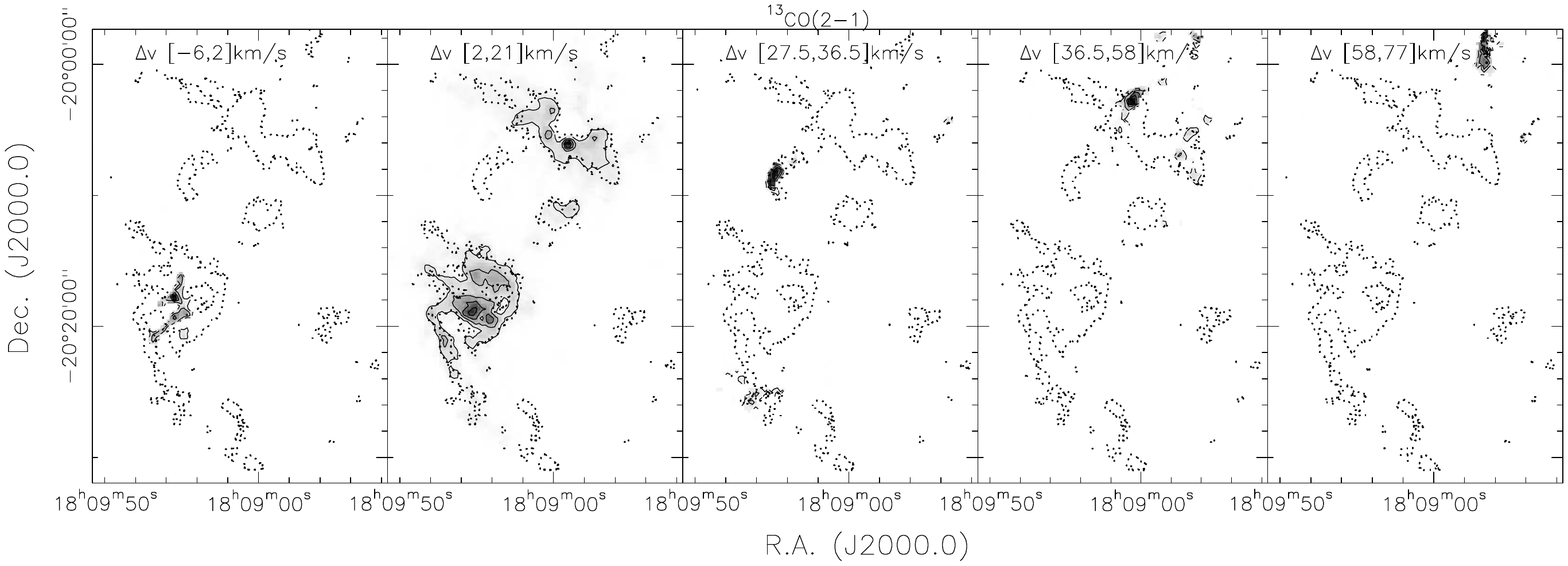}
\caption{The grey-scale shows different $^{13}$CO(2--1) integrated
  intensity images where the integration regimes are shown in the top
  of each panel. Contour levels go from 20 to 80\% in 20\% levels of
  the respective peak intensities. The dotted contours outline
  the 875\,$\mu$m emission at a $3\sigma$ level from
  Fig.~\ref{continuum}.}
\label{13comean}
\end{figure*}

\begin{figure*}[htb]
\includegraphics[angle=-90,width=210mm]{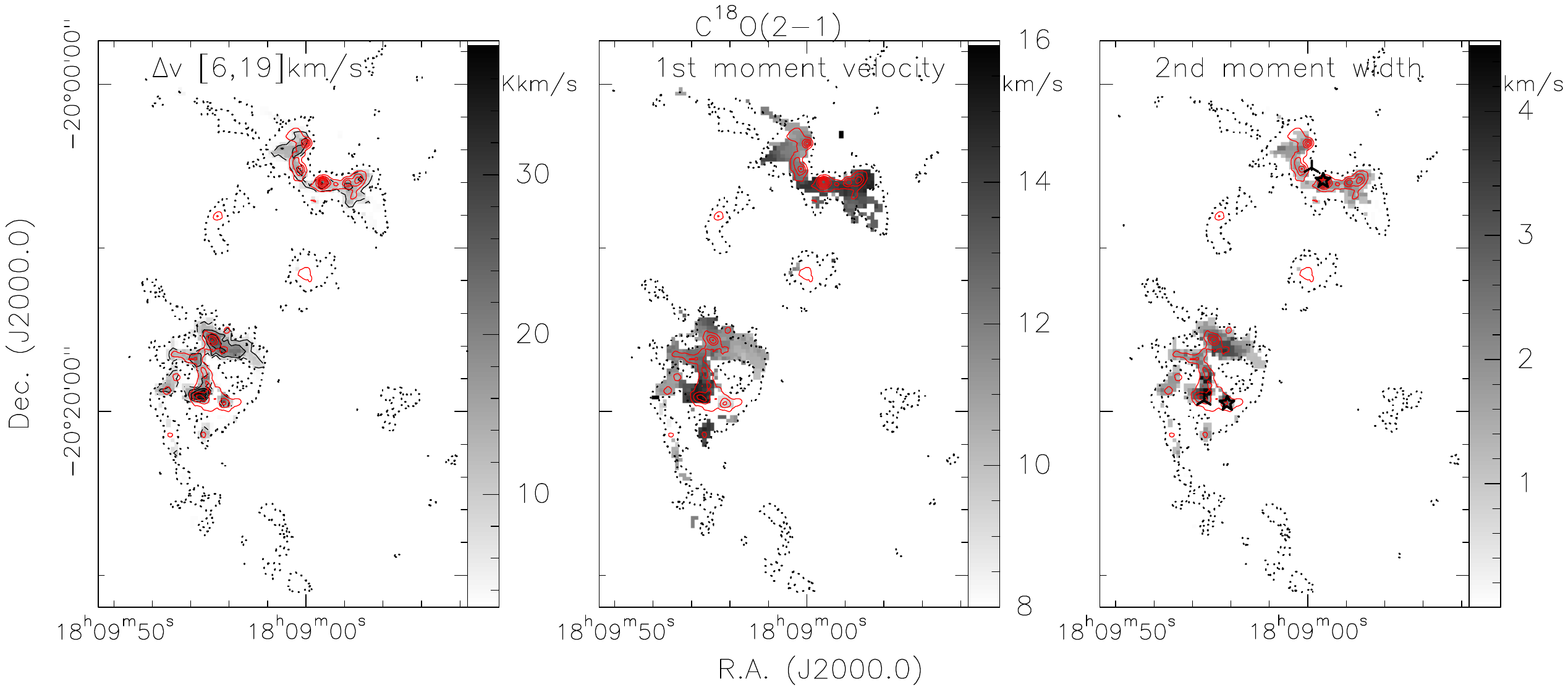}
\caption{The grey-scale shows C$^{18}$O(2--1) moment maps: 0th moment
  or integrated intensity on the left, 1st moment or
  intensity-weighted velocity in the middle and 2nd moment or
  intensity-weighted line-width to the right. The black contours in
  the left panel show the integrated C$^{18}$O(2--1) data and go from
  20 to 80\% in 20\% steps. The dotted contours outline the
  875\,$\mu$m emission at a $3\sigma$ level from Fig.~\ref{continuum},
  and the red contours show additionally the high-intensity
  875\,$\mu$m emission in 15$\sigma$ steps from 15$\sigma$ onwards
  ($1\sigma\approx 70$\,mJy\,beam$^{-1}$). In the right panel, the
  3-pointed stars in the north and south mark the position of an
  O-star by \citet{bik2005} and the center of the southern cluster
  discussed by \citet{blum2001}. The 5-pointed stars mark the two
  UCH{\sc ii} regions \citep{wc1989b}.}
\label{c18o}
\end{figure*}

Figure \ref{13comean} presents $^{13}$CO(2--1) integrated intensity
images of different parts of the $^{13}$CO(2--1) spectrum from Figure
\ref{spectra}. We omit the velocity structure lower than
-6\,km\,s$^{-1}$ because this is only lower-intensity more diffuse
emission which is hard to properly image. However, it very likely
belongs to the overall W31 region. The main velocity component present
in both W31 complexes (G10.2-0.3 and G10.3-0.1) is the main peak
between 2 and 21\,km\,s$^{-1}$. In contrast to this, the two velocity
components between -6 and 2\,km\,s$^{-1}$ and 36.5 to 58\,km\,s$^{-1}$
are clearly only associated with specific sub-regions within G10.2-0.3
and G10.3-0.1, respectively. As outlined in the previous section, the
fact that we see the feature between 27.5 and 36.5\,km\,s$^{-1}$ not
at any other velocity but only as a negative feature throughout the
map due to emission from the OFF-position supports the idea that this
structure is likely associated with other clouds outside our field of
view. Similarly, the structure toward the north between 58 and
77\,km\,s$^{-1}$ may also not be associated with the W31 region but
could be at a different distance. While the main component between 2
and 21\,km\,s$^{-1}$ is widely distributed, it is difficult to safely
distinguish whether the other velocity components in the region are
simply chance alignments of projected sources at different distances,
or whether there exists an additional (projected) large-scale velocity
gradient from blue-shifted emission toward the south-east to more
red-shifted emission toward the north-west. As visible in
Fig.~\ref{spectra}, the averaged C$^{18}$O(2--1) spectrum exhibits in
general similar velocity features as that of $^{13}$CO(2--1), however,
at a lower level. Therefore, in Fig.~\ref{c18o} we only show the
integrated C$^{18}$O(2--1) emission of the main velocity component
between 6 and 19\,km\,s$^{-1}$.

\begin{figure*}[htb]
\includegraphics[angle=-90,width=210mm]{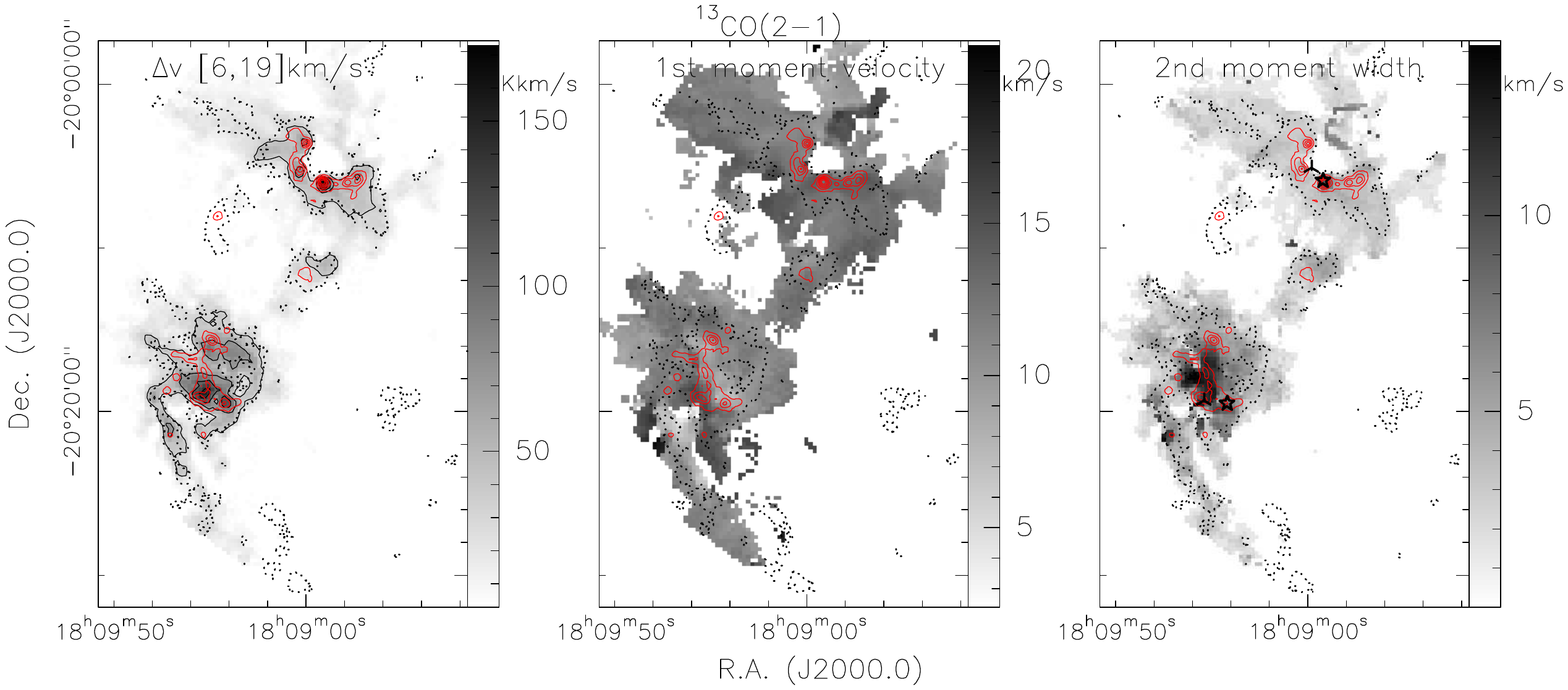}
\caption{The grey-scale shows different $^{13}$CO(2--1) moment maps:
  0th moment or integrated intensity on the left, 1st moment or
  intensity-weighted velocity in the middle and 2nd moment or
  intensity-weighted line-width to the right. The black contours in
  the left panel show the integrated intensity $^{13}$CO(2--1) data
  from 20 to 80\% in 20\% steps. The dotted contours outline the
  875\,$\mu$m emission at a $3\sigma$ level from Fig.~\ref{continuum},
  and the red contours show additionally the high-intensity
  875\,$\mu$m emission in 15$\sigma$ steps from 15$\sigma$ onwards
  ($1\sigma\approx 70$\,mJy\,beam$^{-1}$). In the right panel, the
  3-pointed stars in the north and south mark the position of an
  O-star as reported by \citet{bik2005} and the center of the southern
  cluster discussed by \citet{blum2001}. The 5-pointed stars mark the
  two UCH{\sc ii} regions \citep{wc1989b}.}
\label{13co}
\end{figure*}

Figures \ref{c18o} and \ref{13co} present the 0th, 1st and 2nd moment
maps (integrated intensity, intensity-weighted velocity structure and
intensity-weighted line-width structure) of the main velocity
components of C$^{18}$O(2--1) and $^{13}$CO(2--1), respectively. While
the general structure of the $^{13}$CO and C$^{18}$O emission of this
main velocity components agrees relatively well with the 875\,$\mu$m
continuum emission, in particular the integrated $^{13}$CO(2--1) maps
exhibits much more extended gas structure than the dust continuum
emission. This is likely a combined effect of the limited sensitivity
of the ATLASGAL data on the one hand (approximate rms of
50\,mJy\,beam$^{-1}$, \citealt{schuller2009}), and on the other hand
it may also be due to the general large-scale spatial filtering effect
inherent to any bolometer observations. Furthermore, the C$^{18}$O
integrated emission spatial structure agrees well with the high column
density part of the dust continuum maps outlined by the red contours
in Figure \ref{c18o}.

While some sub-regions exhibit additional velocity components (see
Fig.~\ref{13comean} and discussion above), the 1st moment maps in
Figs.~\ref{c18o} and \ref{13co} do not exhibit a clear velocity
difference between the two large-scale sub-complexes within the main
velocity component. Rather in contrast, both regions -- G10.2-0.3 and
G10.3-0.1 -- show velocity sub-structure associated with the width of
the main velocity component. Probably more important, the 2nd moment
maps show a very broad distribution of line-widths, extending in
$^{13}$CO(2--1) to values in excess of 10\,km\,s$^{-1}$. Although
still comparably narrow (see $\Delta v$ for clump 1 in Tables
\ref{clump_parameters} \& \ref{clump_parameters_smooth}) in the
northern region (G10.3-0.1), the broadest line-width there is measured
toward clump 1 (Table \ref{clump_parameters_smooth}), which is
spatially associated with the near-infrared cluster around the O-stars
\citep{bik2005}, hence the most evolved part of this sub-complex.
Although the southern region G10.2-0.3 exhibits in general much
broader line widths compared to the northern region, there the
broadest C$^{18}$O(2--1) and $^{13}$CO(2--1) line-widths are
attributed again to regions in the close vicinity of the infrared
cluster discussed by \citet{blum2001}. Toward the southern and
northern regions, the UCH{\sc ii} regions both exhibit comparably
broad line width, however, always narrower than toward the gas clumps
associated with the infrared clusters.

\subsection{Line parameters and virial analysis}
\label{virial}

In particular the C$^{18}$O(2--1) spectral line data are suited for a
more detailed spectral analysis. Therefore, we extracted the
C$^{18}$O(2--1) spectra toward all peak positions identified by the
clumpfind procedure for the dust continuum data. However, one should
keep in mind that the spatial resolution of the continuum and line
data is different, $19.2''$ and $27.5''$, respectively. While for
completeness, we extract the spectra toward all positions from the
high-spatial-resolution continuum data (Table \ref{clump_parameters}),
for a better comparison, it is more useful to extract the line
parameters toward the positions extracted from the smoothed
875\,$\mu$m continuum map (Section \ref{smooth}). Table
\ref{clump_parameters_smooth} lists the derived line parameters toward
the smoothed dust continuum peak positions from Gaussian fits to the
C$^{18}$O(2--1) line profiles. Line parameters are peak temperature
$T_{\rm{peak}}$, integrated intensity $\int T_{\rm{peak}} dv$, peak
velocity $v_{\rm{peak}}$ and Full Width Half Maximum line width
$\Delta v$.

\begin{figure}[htb]
\includegraphics[angle=-90,width=90mm]{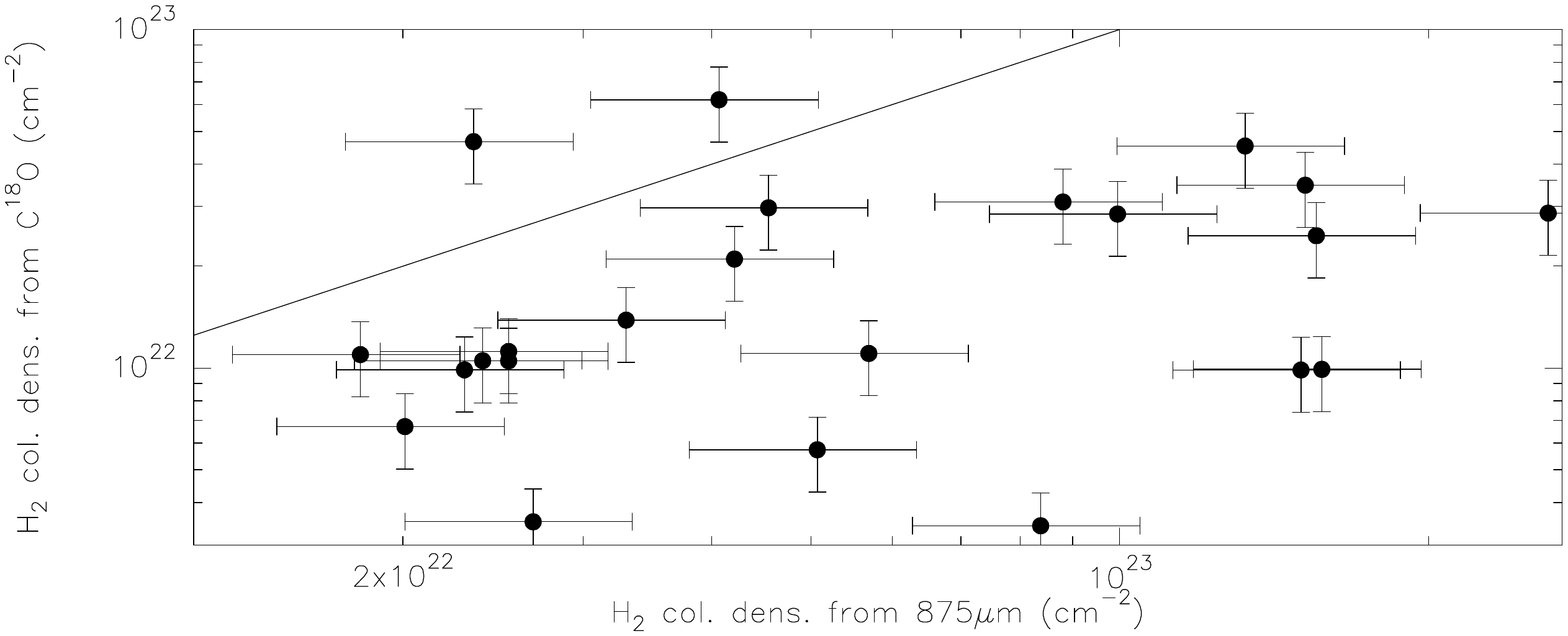}
\caption{Comparison of H$_2$ column densities derived from the
  875\,$\mu$m dust continuum and C$^{18}$O(2-1) at the same spatial
  resolution of $27.5''$. We draw 25\% error-bars although the real
  errors may be even larger as discussed in the main text. The solid
  line marks the 1:1 relation.}
\label{column}
\end{figure}

These line parameters can be used to derive physical quantities, in
particular H$_2$ column densities and virial masses. To calculate the
C$^{18}$O and corresponding H$_2$ column densities, we followed the
standard Local Thermodynamic Equilibrium (LTE) analysis
\citep{rohlfs2006} assuming optically thin C$^{18}$O(2--1) emission
again at an average temperature of 50\,K (section \ref{clump_prop}).
Comparing individual $^{13}$CO and C$^{18}$O spectra of the region
indicates that this assumption is reasonable. The resulting H$_2$
column densities are listed in Tables \ref{clump_parameters} and
\ref{clump_parameters_smooth}. While higher/lower temperatures for
individual sub-sources would lower/raise the column density estimates,
we use again a uniform temperature to better compare later with the
dust continuum derived results (Fig.~\ref{column}). Non-optically thin
C$^{18}$O emission would also raise the derived column density values.

Furthermore, we calculated virial masses for the given C$^{18}$O(2--1)
line widths following \citet{maclaren1988}

\begin{eqnarray}
M_{\rm{vir}} = k_2\, R\, \Delta v^2
\label{virial_eq}
\end{eqnarray}

with $k_2=126$ for a density profile $\rho\propto r^{-2}$, $R$ the
radius of the clump defined as half of the FWHM of the beam and
$\Delta v$ the line width given in Table
\ref{clump_parameters_smooth}.  Under the assumption of a flatter
density profile ($\rho\propto r^{-1}$) or a Gaussian density
distribution the virial masses would be approximately factors of 1.5
or 3 higher \citep{maclaren1988,simon2001}. We assume a factor 2
uncertainty for the calculated virial masses. The derived masses are
listed also in Tables \ref{clump_parameters} and
\ref{clump_parameters_smooth}.

Figure \ref{column} presents a comparison of the H$_2$ column
densities derived on the one hand via the 875\,$\mu$m continuum
emission and on the other hand from the C$^{18}$O(2--1) emission at
the same spatial resolution of $27.5''$. Although the real errors may
even be larger (see discussions above), for clarity we draw ``only''
25\% error margins in the figure. For almost all positions, the H$_2$
column densities derived from the 875\,$\mu$m data exceed those
derived via the C$^{18}$O emission. Even considering the given
uncertainties for the different column density derivations, this is
plausible since with a critical density of $\sim 10^4$\,cm$^{-3}$,
C$^{18}$O(2--1) does not trace the highest density gas and therefore
may miss the highest column density regions.  Similar results were
recently obtained by \citet{walsh2010} for the regions NGC6334I \&
I(N) where even the very optically thin CO isotopologue
C$^{17}$O(1--0) did not trace the dust continuum derived column
density peak of the region.

\begin{figure}[htb]
\includegraphics[angle=-90,width=90mm]{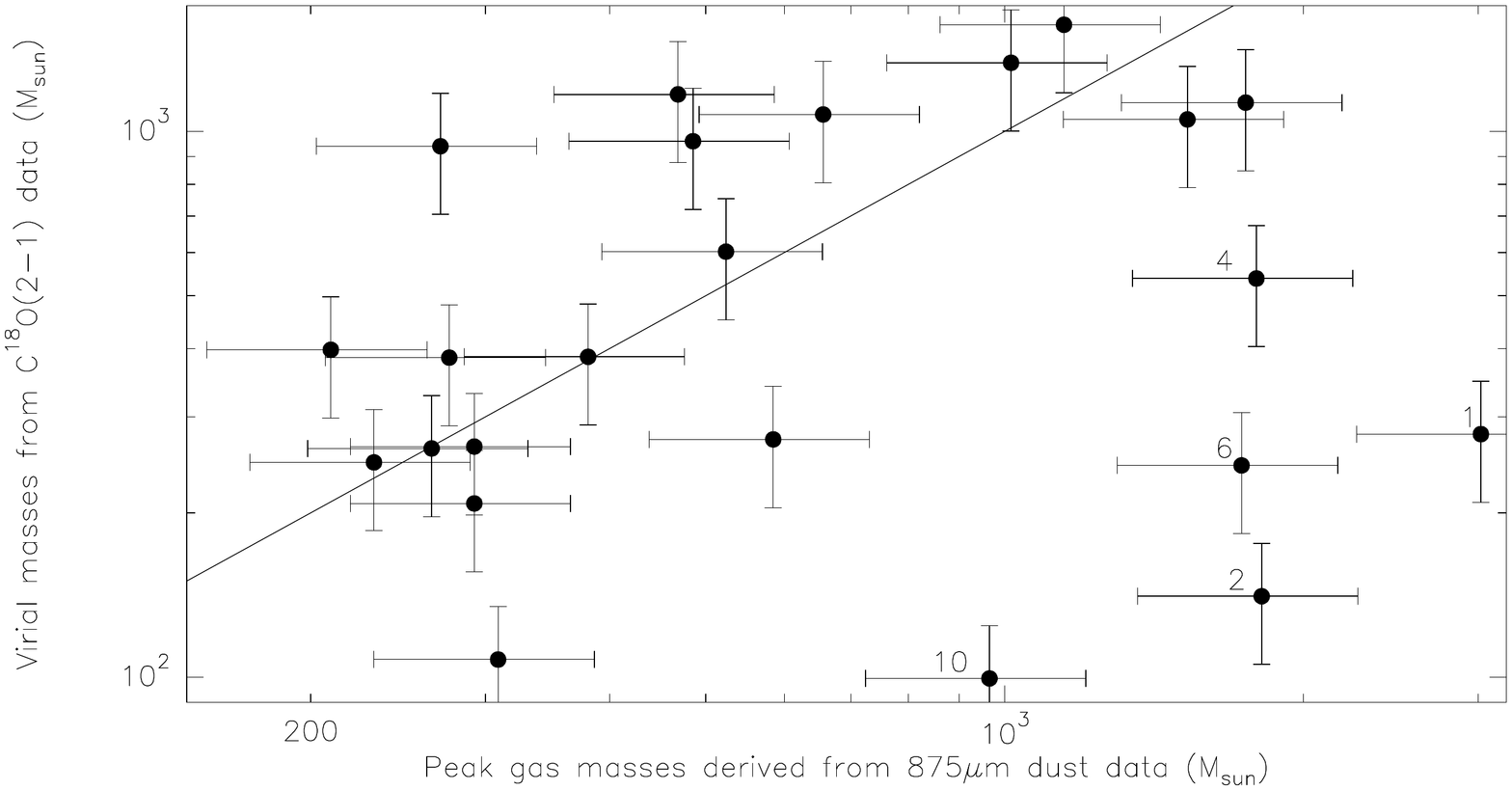}
\caption{Comparison of masses derived from the 875\,$\mu$m dust
  continuum as well as via the C$^{18}$O(2-1) line width under the
  assumption of virial equilibrium at the same spatial resolution of
  $27.5''$. We draw 25\% error-bars although the real errors may be
  even larger as discussed in the main text. The solid line marks the
  1:1 relation. The numbers in the bottom-right part correspond to the
  clumps discussed in the main text and marked in Figure
  \ref{clumps}.}
\label{masses}
\end{figure}

A different comparison is presented in Figure \ref{masses} where we
show the gas masses derived from the 875\,$\mu$m dust continuum peak
fluxes $M_{\rm{peak}}$ against the masses obtained from the
C$^{18}$O(2--1) line width under the assumption of virial equilibrium
M$_{\rm{vir}}$. Similarly to the column density figure, we draw
``only'' 25\% error margins for clarity reasons. To first order, we
find no clear trend but a ``scatter plot''.  However, excluding clumps
1, 2, 4, 6 and 10 (Fig.~\ref{clumps} and Table
\ref{clump_parameters_smooth}), although with a considerable scatter
most other clumps are located in the vicinity of the
$M_{\rm{vir}}/M_{\rm{peak}}\sim 1$ relation. Therefore, many of these
clumps may not be far from virial equilibrium. Inspecting clumps 1, 2,
4, 6 and 10 below the 1:1 relation in Figure \ref{masses} in more
detail, interestingly we find that all these sources are associated
with the northern sub-region G10.3-0.1 (Figs~\ref{masses},
\ref{clumps} \& \ref{zoom}), whereas the southern region G10.2-0.3 is
dominated by higher $M_{\rm{vir}}/M_{\rm{peak}}$ ratios. Hence, there
appears to be a significant difference in the turbulent line width
contribution between both sub-regions (see discussion in section
\ref{evolution}).

\section{Discussion}

\subsection{Different evolutionary stages and kinematic properties}
\label{evolution}

As already mentioned in the introduction, the spatial offset between
near-infrared clusters and UCH{\sc ii} regions toward the northern
G10.3-0.1 and the southern G10.2-0.3 complex are indicative of several
episodes of high-mass star formation within each of the sub-regions.
Furthermore, as outlined in section \ref{general} and Fig.~\ref{zoom},
within the G10.3-0.1 region, we do not only have the infrared cluster
and the UCH{\sc ii} region, but we also find at least one high-mass gas
clump G10.3W2 without any cm or mid-infrared emission indicative of
star formation. Hence this clump may still be in a starless phase
prior to active star formation. Therefore, from an evolutionary point
of view, the northern G10.3-0.1 complex hosts several high-mass
star-forming gas and dust clumps -- all potentially capable to form
high-mass clusters -- in at least three different evolutionary stages: a
more evolved infrared cluster where the main sequence stars have
already emerged, high-mass protostellar objects with and without
embedded UCH{\sc ii} regions that may still be in an ongoing accretion
phase, and high-mass starless clump candidates (see Table
\ref{clump_parameters} for their parameters). Although there is no
robust proof that the infrared cluster actually triggered the star
formation processes in the neighboring clumps, the spatial structure
with the infrared cluster almost at the geometrical center of
G10.3-0.1, then the high-mass protostellar objects following in the
surrounding, and the high-mass starless clump candidates at the edge
of the region, indicates that triggered star formation may be a
possible cause for the different evolutionary stages in this region.

Toward the southern G10.2-0.3 complex we also identify an infrared
cluster and several high-mass protostellar objects, but there we do
not as unambiguously find regions classifying as high-mass starless
clump candidates. For a broader discussion of evolutionary sequences
in high-mass star formation, we refer to the reviews by
\citet{beuther2006b} and \citet{zinnecker2007}.

In addition to the evolutionary differences within each of the two
main regions in W31, the kinematic analysis in section \ref{virial} is
also indicative of evolutionary differences between the two complexes
in general. The virial mass as presented in section \ref{virial} is
directly proportional to the observed line width
(eq.~\ref{virial_eq}). In the simple virial equilibrium picture,
having a ratio $M_{\rm{vir}}/M_{\rm{peak}}\sim 1$ would imply that the
clumps could be stable against collapse. However, except for a few
notable sources like G10.3W2 (Fig.~\ref{zoom} and clump 6 in
Fig.~\ref{clumps}, see also Table \ref{clump_parameters_smooth} and
Fig.~\ref{masses}) which are candidates for starless clumps, most
other regions in this complex show various signs of star formation.
Hence, they are unlikely to be in virial equilibrium.  It is rather
the opposite, the feedback from ongoing star formation (e.g., outflows
or radiation) significantly broadens the observed line width and the
virial assumption is not properly applicable under such circumstances.

While we do not know whether the clumps with high
$M_{\rm{vir}}/M_{\rm{peak}}$ ratios are still bound or already
expanding again from the inner energy sources, the fact that we see
ongoing star formation indicates that they are still at least partly
bound. Large virial parameters are also consistent with
pressure-confined clumps as discussed by \citet{bertoldi1992} and more
recently by \citet{lada2008}, \citet{dobbs2011} and Kainulainen et
al.~(subm.). High virial mass ratios have also been found by
\citet{simon2001} for four molecular clouds studied via the Galactic
Ring Survey (GRS, \citealt{jackson2006}). It is interesting to note
that the three less evolved clouds of their sample have on average
even higher virial mass ratios than their most luminous source, the
galactic mini-starburst W49. The peak of their ratio between virial
mass and LTE mass (derived via LTE calculations for the
$^{13}$CO(1--0) emission) for W49 is around 2, whereas we find an
average of $M_{\rm{vir}}/M_{\rm{peak}}$ for W31 of $\sim 1.1$. Hence,
similar to W49 studied by \citet{simon2001}, the clumps in W31 are
in the regime of being bound.

Maybe more surprising is that all high-mass clumps associated with
G10.3-0.1 have $M_{\rm{vir}}/M_{\rm{peak}}$ ratios below 1. While this
could be expected for the starless clump candidate G10.3W2 if it were
in an unstable state just at the verge of collapse and had no
additional turbulent support, it is less expected for the other active
star-forming clumps where multiple additional line width broadening
mechanisms should be at work, e.g., outflows and/or radiation.
Nevertheless, low $M_{\rm{vir}}/M_{\rm{peak}}$ ratios clearly imply a
comparatively small turbulent line width with respect to most of the
clumps associated with the southern region G10.2-0.3.  Furthermore,
magnetic fields may be more important in these supposedly younger
clumps acting as additional source of stability, also implying low
$M_{\rm{vir}}/M_{\rm{peak}}$ ratios.

While the overall characteristics of the two complexes G10.2-0.3 and
G10.3-0.1 appear similar -- both have similar luminosities (see
section \ref{intro}), strong submm and mid-infrared emission as well
as an associated near-infrared cluster -- there is an apparent
evolutionary difference between the two. In addition to the above
discussed different $M_{\rm{vir}}/M_{\rm{peak}}$ ratios between
G10.2-0.3 and G10.3-0.1, the class 0/I and class II sources around the
southern G10.2-0.3 region appear considerably more widely
distributed than those around the northern region G10.3-0.1 (see
section \ref{general}). Furthermore, it is also interesting to note
that only G10.3-0.1 exhibits several sites of Class {\sc ii} CH$_3$OH
masers whereas none is found toward G10.2-0.3 \citep{walsh1998}. Since
these CH$_3$OH masers are also usually found toward younger regions
(e.g., \citealt{fish2007}), the presence of them toward only the
northern region G10.3-0.1 combined with the above discussed additional
differences strongly suggests that the northern G10.3-0.1 complex on
average should be younger than the southern G10.2-0.3 region. The
latter then had more time already to stir up the gas of the
environment and hence increase the observed line width.

On top of all these evolutionary effects, we also find a closer
association of the IRAC-identified younger class 0/I sources with the
central high-mass gas clumps while the more evolved class II sources
appear more broadly distributed over the whole field
(Fig.~\ref{continuum}).

\subsection{The shape of the clump mass function}
\label{cmf_discussion}

The power-law index $\alpha \sim 1.5\pm 0.3$ of the clump mass
function discussed in Fig.~\ref{cmf} ($\Delta N/\Delta M \propto
M^{-\alpha}$) is consistent with the clump mass function derived for
molecular clouds by molecular line CO observations ($\alpha\sim 1.6$,
e.g., \citealt{stutzki1990,blitz1993,kramer1998,simon2001}), but it is
considerably flatter than distributions derived for other star-forming
regions from dust continuum observations or extinction mapping, which
more resemble the Salpeter slope ($\alpha\sim 2.35$, e.g.,
\citealt{motte1998,beltran2006,reid2006,alves2007}). In the past, the
differences in power-law mass distributions derived from the different
tracers (CO versus dust emission/extinction) was often attributed to
the different density regimes they are tracing: The CO observations
are more sensitive to the diffuse and transient gas whereas the dust
emission/extinction rather traces the denser, gravitationally or
pressure-bound cores. In this picture, the different power-law
distributions would reflect a structural change of the cloud/clump
properties from transient to bound structures.

However, this picture does not hold for our data of the W31 complex
since we know that the region is actively star-forming and not a
transient structure. Since cluster mass functions are also flatter
than stellar mass functions (power-law distributions of 2.0, e.g.,
\citealt{bik2003} or \citealt{degrijs2003}, versus 2.35 of the
Salpeter slope, respectively), one possibility to explain the
discrepancies is that we are tracing high-mass cluster-forming regions
at far distances like W31 whereas the observations by
\citet{motte1998} or \citet{alves2007} dissect low-mass regions like
$\rho$ Ophiuchus or the Pipe at distances closer $\leq 150$\,pc.
Unfortunately, this explanation is not consistent with all
observations since the mass distributions derived by
\citet{beltran2006} or \citet{reid2006} are also targeting high-mass
star-forming regions at distances of several kpc.

Another possibility for a flatter clump mass function with respect to
the IMF could be that the star formation efficiency (SFE) varies with
clump mass. Higher mass gas clumps may exhibit on average a
lower SFE which would steepen the slope from the clump mass function
to the IMF.  Similar results were recently inferred by
\citet{parmentier2011}. In a similar direction, it is worth noting
that \citet{simon2001} also inferred for their most luminous
mini-starburst W49 the flattest clump mass function with a power-law
index of 1.56, in very close agreement with our result for the
high-luminosity region W31. Therefore, there are several indications
that the most luminous regions could also exhibit the flattest clump
mass functions.

One potentially more technical explanation for the relatively flat
clump mass function we derive for W31 could be that our used uniform
temperature of 50\,K more severely affects $\alpha$ than we
anticipated. Higher temperatures decrease the mass estimates of a
clump whereas lower temperatures raise the mass estimates. In this
picture, it could well be that the higher mass clumps closer to the
centers of the region have higher temperatures whereas lower mass
clumps may still be colder. This effect would steepen the mass
function. Although with the current data we cannot properly quantify
this effect, it is unlikely that the mass function would steepen to a
Salpeter slope. 

Therefore, we suggest that the observed flat mass function reflects
the facts that the resulting cluster mass functions are also flatter
than the stellar initial mass function, and furthermore that the SFE
may vary with clump mass.

\section{Conclusions}

The most luminous and high-mass star formation appears to take place in
a highly structured and non-uniform fashion. The multi-wavelength
continuum and spectral line analysis of the different sub-regions
within the very luminous W31 regions reveals evolutionary differences
on large scales between the two sub-complexes G10.2-0.3 and G10.3-0.1,
but also within each of these regions on much smaller scales.  While
many clumps have virial mass ratios close to unity, we also find
low-turbulence, potentially still starless high-mass gas clumps that
can reside in almost the direct vicinity of ongoing and already
finished high-mass cluster-forming regions. Furthermore, these dense
active gas clumps appear to be surrounded by a halo-like distribution
of already more evolved class 0/1 and class II sources. We also find a
tight correlation between the warm dust tracing 24\,$\mu$m emission
and the ionized gas tracing cm emission, implying that warm dust at
temperatures around 1000\,K can spatially coexist with ionized gas in
the $10^4$\,K regime. In addition to this, our data indicate that the
clump mass function of W31 is considerably flatter (power-law index
$\alpha \sim 1.5\pm 0.3$) than the initial mass function, but it is
consistent with the mass functions derived for other molecular clouds.
Since these gas clumps trace cluster scales, this is consistent with
the flatter cluster mass function compared to the stellar mass
function.  Furthermore, the analysis tentatively suggests that the
star formation efficiency may decrease with increasing clump mass.



\begin{thebibliography}{68}
\expandafter\ifx\csname natexlab\endcsname\relax\def\natexlab#1{#1}\fi

\bibitem[{{Allen} {et~al.}(2004){Allen}, {Calvet}, {D'Alessio}, {Merin},
  {Hartmann}, {Megeath}, {Gutermuth}, {Muzerolle}, {Pipher}, {Myers}, \&
  {Fazio}}]{allen2004}
{Allen}, L.~E., {Calvet}, N., {D'Alessio}, P., {et~al.} 2004, \apjs, 154, 363

\bibitem[{{Alves} {et~al.}(2007){Alves}, {Lombardi}, \& {Lada}}]{alves2007}
{Alves}, J., {Lombardi}, M., \& {Lada}, C.~J. 2007, \aap, 462, L17

\bibitem[{{Beltr{\'a}n} {et~al.}(2006){Beltr{\'a}n}, {Brand}, {Cesaroni},
  {Fontani}, {Pezzuto}, {Testi}, \& {Molinari}}]{beltran2006}
{Beltr{\'a}n}, M.~T., {Brand}, J., {Cesaroni}, R., {et~al.} 2006, \aap, 447,
  221

\bibitem[{{Bertoldi} \& {McKee}(1992)}]{bertoldi1992}
{Bertoldi}, F. \& {McKee}, C.~F. 1992, \apj, 395, 140

\bibitem[{{Beuther} {et~al.}(2007){Beuther}, {Churchwell}, {McKee}, \&
  {Tan}}]{beuther2006b}
{Beuther}, H., {Churchwell}, E.~B., {McKee}, C.~F., \& {Tan}, J.~C. 2007, in
  Protostars and Planets V, ed. B.~{Reipurth}, D.~{Jewitt}, \& K.~{Keil},
  165--180

\bibitem[{{Beuther} {et~al.}(2002){Beuther}, {Schilke}, {Menten}, {Motte},
  {Sridharan}, \& {Wyrowski}}]{beuther2002a}
{Beuther}, H., {Schilke}, P., {Menten}, K.~M., {et~al.} 2002, \apj, 566, 945

\bibitem[{{Beuther} {et~al.}(2005){Beuther}, {Schilke}, {Menten}, {Motte},
  {Sridharan}, \& {Wyrowski}}]{beuther2002erratum}
{Beuther}, H., {Schilke}, P., {Menten}, K.~M., {et~al.} 2005, \apj, 633, 535

\bibitem[{{Bik}(2004)}]{bikphd}
{Bik}, A. 2004, Ph.D.~Thesis

\bibitem[{{Bik} {et~al.}(2005){Bik}, {Kaper}, {Hanson}, \& {Smits}}]{bik2005}
{Bik}, A., {Kaper}, L., {Hanson}, M.~M., \& {Smits}, M. 2005, \aap, 440, 121

\bibitem[{{Bik} {et~al.}(2003){Bik}, {Lamers}, {Bastian}, {Panagia}, \&
  {Romaniello}}]{bik2003}
{Bik}, A., {Lamers}, H.~J.~G.~L.~M., {Bastian}, N., {Panagia}, N., \&
  {Romaniello}, M. 2003, \aap, 397, 473

\bibitem[{{Blitz}(1993)}]{blitz1993}
{Blitz}, L. 1993, in Protostars and Planets III, 125--161

\bibitem[{{Blum} {et~al.}(2001){Blum}, {Damineli}, \& {Conti}}]{blum2001}
{Blum}, R.~D., {Damineli}, A., \& {Conti}, P.~S. 2001, \aj, 121, 3149

\bibitem[{{Bonnell} {et~al.}(2007){Bonnell}, {Larson}, \&
  {Zinnecker}}]{bonnell2006}
{Bonnell}, I.~A., {Larson}, R.~B., \& {Zinnecker}, H. 2007, in Protostars and
  Planets V, ed. B.~{Reipurth}, D.~{Jewitt}, \& K.~{Keil}, 149--164

\bibitem[{{Carey} {et~al.}(2009){Carey}, {Noriega-Crespo}, {Mizuno}, {Shenoy},
  {Paladini}, {Kraemer}, {Price}, {Flagey}, {Ryan}, {Ingalls}, {Kuchar},
  {Pinheiro Gon{\c c}alves}, {Indebetouw}, {Billot}, {Marleau}, {Padgett},
  {Rebull}, {Bressert}, {Ali}, {Molinari}, {Martin}, {Berriman}, {Boulanger},
  {Latter}, {Miville-Deschenes}, {Shipman}, \& {Testi}}]{carey2009}
{Carey}, S.~J., {Noriega-Crespo}, A., {Mizuno}, D.~R., {et~al.} 2009, \pasp,
  121, 76

\bibitem[{{Cesaroni} {et~al.}(1994){Cesaroni}, {Churchwell}, {Hofner},
  {Walmsley}, \& {Kurtz}}]{cesaroni1994}
{Cesaroni}, R., {Churchwell}, E., {Hofner}, P., {Walmsley}, C.~M., \& {Kurtz},
  S. 1994, \aap, 288, 903

\bibitem[{{Churchwell} {et~al.}(2009){Churchwell}, {Babler}, {Meade},
  {Whitney}, {Benjamin}, {Indebetouw}, {Cyganowski}, {Robitaille}, {Povich},
  {Watson}, \& {Bracker}}]{churchwell2009}
{Churchwell}, E., {Babler}, B.~L., {Meade}, M.~R., {et~al.} 2009, \pasp, 121,
  213

\bibitem[{{Corbel} {et~al.}(1997){Corbel}, {Wallyn}, {Dame}, {Durouchoux},
  {Mahoney}, {Vilhu}, \& {Grindlay}}]{corbel1997}
{Corbel}, S., {Wallyn}, P., {Dame}, T.~M., {et~al.} 1997, \apj, 478, 624

\bibitem[{{de Grijs} {et~al.}(2003){de Grijs}, {Anders}, {Bastian}, {Lynds},
  {Lamers}, \& {O'Neil}}]{degrijs2003}
{de Grijs}, R., {Anders}, P., {Bastian}, N., {et~al.} 2003, \mnras, 343, 1285

\bibitem[{{Dobbs} {et~al.}(2011){Dobbs}, {Burkert}, \& {Pringle}}]{dobbs2011}
{Dobbs}, C.~L., {Burkert}, A., \& {Pringle}, J.~E. 2011, ArXiv e-prints

\bibitem[{{Downes} {et~al.}(1980){Downes}, {Wilson}, {Bieging}, \&
  {Wink}}]{downes1980}
{Downes}, D., {Wilson}, T.~L., {Bieging}, J., \& {Wink}, J. 1980, \aaps, 40,
  379

\bibitem[{{Draine}(2003)}]{draine2003}
{Draine}, B.~T. 2003, \araa, 41, 241

\bibitem[{{Draine} {et~al.}(2007){Draine}, {Dale}, {Bendo}, {Gordon}, {Smith},
  {Armus}, {Engelbracht}, {Helou}, {Kennicutt}, {Li}, {Roussel}, {Walter},
  {Calzetti}, {Moustakas}, {Murphy}, {Rieke}, {Bot}, {Hollenbach}, {Sheth}, \&
  {Teplitz}}]{draine2007}
{Draine}, B.~T., {Dale}, D.~A., {Bendo}, G., {et~al.} 2007, \apj, 663, 866

\bibitem[{{Everett} \& {Churchwell}(2010)}]{everett2010}
{Everett}, J.~E. \& {Churchwell}, E. 2010, \apj, 713, 592

\bibitem[{{Fish}(2007)}]{fish2007}
{Fish}, V.~L. 2007, in IAU Symposium, Vol. 242, IAU Symposium, ed.
  {J.~M.~Chapman \& W.~A.~Baan}, 71--80

\bibitem[{{Fontani} {et~al.}(2005){Fontani}, {Beltr{\'a}n}, {Brand},
  {Cesaroni}, {Testi}, {Molinari}, \& {Walmsley}}]{fontani2005}
{Fontani}, F., {Beltr{\'a}n}, M.~T., {Brand}, J., {et~al.} 2005, \aap, 432, 921

\bibitem[{{Ghosh} {et~al.}(1989){Ghosh}, {Iyengar}, {Rengarajan}, {Tandon},
  {Verma}, {Daniel}, \& {Ho}}]{ghosh1989}
{Ghosh}, S.~K., {Iyengar}, K.~V.~K., {Rengarajan}, T.~N., {et~al.} 1989, \apj,
  347, 338

\bibitem[{{G{\"u}sten} {et~al.}(2006){G{\"u}sten}, {Nyman}, {Schilke},
  {Menten}, {Cesarsky}, \& {Booth}}]{guesten2006}
{G{\"u}sten}, R., {Nyman}, L.~{\AA}., {Schilke}, P., {et~al.} 2006, \aap, 454,
  L13

\bibitem[{{Helfand} {et~al.}(2006){Helfand}, {Becker}, {White}, {Fallon}, \&
  {Tuttle}}]{helfand2006}
{Helfand}, D.~J., {Becker}, R.~H., {White}, R.~L., {Fallon}, A., \& {Tuttle},
  S. 2006, \aj, 131, 2525

\bibitem[{{Hildebrand}(1983)}]{hildebrand1983}
{Hildebrand}, R.~H. 1983, \qjras, 24, 267

\bibitem[{{Homeier} \& {Alves}(2005)}]{homeier2005}
{Homeier}, N.~L. \& {Alves}, J. 2005, \aap, 430, 481

\bibitem[{{Jackson} {et~al.}(2006){Jackson}, {Rathborne}, {Shah}, {Simon},
  {Bania}, {Clemens}, {Chambers}, {Johnson}, {Dormody}, {Lavoie}, \&
  {Heyer}}]{jackson2006}
{Jackson}, J.~M., {Rathborne}, J.~M., {Shah}, R.~Y., {et~al.} 2006, \apjs, 163,
  145

\bibitem[{{Jenkins}(2004)}]{jenkins2004}
{Jenkins}, E.~B. 2004, in Origin and Evolution of the Elements, ed.
  A.~{McWilliam} \& M.~{Rauch}, 336

\bibitem[{{Johnston} {et~al.}(2009){Johnston}, {Shepherd}, {Aguirre}, {Dunham},
  {Rosolowsky}, \& {Wood}}]{johnston2009}
{Johnston}, K.~G., {Shepherd}, D.~S., {Aguirre}, J.~E., {et~al.} 2009, \apj,
  707, 283

\bibitem[{{Keto}(2002)}]{keto2002a}
{Keto}, E. 2002, \apj, 568, 754

\bibitem[{{Kim} \& {Koo}(2002)}]{kim2002}
{Kim}, K.-T. \& {Koo}, B.-C. 2002, \apj, 575, 327

\bibitem[{{Klein} {et~al.}(2006){Klein}, {Philipp}, {Kr{\"a}mer}, {Kasemann},
  {G{\"u}sten}, \& {Menten}}]{klein2006}
{Klein}, B., {Philipp}, S.~D., {Kr{\"a}mer}, I., {et~al.} 2006, \aap, 454, L29

\bibitem[{{Kramer} {et~al.}(1998){Kramer}, {Stutzki}, {Rohrig}, \&
  {Corneliussen}}]{kramer1998}
{Kramer}, C., {Stutzki}, J., {Rohrig}, R., \& {Corneliussen}, U. 1998, \aap,
  329, 249

\bibitem[{{Kroupa}(2001)}]{kroupa2001}
{Kroupa}, P. 2001, \mnras, 322, 231

\bibitem[{{Krumholz} \& {Bonnell}(2007)}]{krumholz2008a}
{Krumholz}, M.~R. \& {Bonnell}, I.~A. 2007, ArXiv e-prints,0712.0828

\bibitem[{{Krumholz} \& {McKee}(2008)}]{krumholz2008b}
{Krumholz}, M.~R. \& {McKee}, C.~F. 2008, \nat, 451, 1082

\bibitem[{{Lada} {et~al.}(2008){Lada}, {Muench}, {Rathborne}, {Alves}, \&
  {Lombardi}}]{lada2008}
{Lada}, C.~J., {Muench}, A.~A., {Rathborne}, J., {Alves}, J.~F., \& {Lombardi},
  M. 2008, \apj, 672, 410

\bibitem[{{MacLaren} {et~al.}(1988){MacLaren}, {Richardson}, \&
  {Wolfendale}}]{maclaren1988}
{MacLaren}, I., {Richardson}, K.~M., \& {Wolfendale}, A.~W. 1988, \apj, 333,
  821

\bibitem[{{McKee} \& {Ostriker}(2007)}]{mckee2007}
{McKee}, C.~F. \& {Ostriker}, E.~C. 2007, \araa, 45, 565

\bibitem[{{Megeath} {et~al.}(2004){Megeath}, {Allen}, {Gutermuth}, {Pipher},
  {Myers}, {Calvet}, {Hartmann}, {Muzerolle}, \& {Fazio}}]{megeath2004}
{Megeath}, S.~T., {Allen}, L.~E., {Gutermuth}, R.~A., {et~al.} 2004, \apjs,
  154, 367

\bibitem[{{Molinari} {et~al.}(1996){Molinari}, {Brand}, {Cesaroni}, \&
  {Palla}}]{molinari1996}
{Molinari}, S., {Brand}, J., {Cesaroni}, R., \& {Palla}, F. 1996, \aap, 308,
  573

\bibitem[{{Motte} {et~al.}(1998){Motte}, {Andre}, \& {Neri}}]{motte1998}
{Motte}, F., {Andre}, P., \& {Neri}, R. 1998, \aap, 336, 150

\bibitem[{{Mueller} {et~al.}(2002){Mueller}, {Shirley}, {Evans}, \&
  {Jacobson}}]{mueller2002}
{Mueller}, K.~E., {Shirley}, Y.~L., {Evans}, N.~J., \& {Jacobson}, H.~R. 2002,
  \apjs, 143, 469

\bibitem[{{Parmentier}(2011)}]{parmentier2011}
{Parmentier}, G. 2011, ArXiv e-prints

\bibitem[{{Plume} {et~al.}(1997){Plume}, {Jaffe}, {Evans}, {Martin-Pintado}, \&
  {Gomez-Gonzalez}}]{plume1997}
{Plume}, R., {Jaffe}, D.~T., {Evans}, N.~J., {Martin-Pintado}, J., \&
  {Gomez-Gonzalez}, J. 1997, \apj, 476, 730

\bibitem[{{Qiu} {et~al.}(2008){Qiu}, {Zhang}, {Megeath}, {Gutermuth},
  {Beuther}, {Shepherd}, {Sridharan}, {Testi}, \& {De Pree}}]{qiu2008}
{Qiu}, K., {Zhang}, Q., {Megeath}, S.~T., {et~al.} 2008, \apj, 685, 1005

\bibitem[{{Reid} \& {Wilson}(2006)}]{reid2006}
{Reid}, M.~A. \& {Wilson}, C.~D. 2006, \apj, 650, 970

\bibitem[{{Rohlfs} \& {Wilson}(2006)}]{rohlfs2006}
{Rohlfs}, K. \& {Wilson}, T.~L. 2006, {Tools of radio astronomy} (Tools of
  radio astronomy, 4th rev.~and enl.~ed., by K.~Rohlfs and T.L.~Wilson.~
  Berlin: Springer, 2006)

\bibitem[{{Sanders} {et~al.}(1986){Sanders}, {Clemens}, {Scoville}, \&
  {Solomon}}]{sanders1986}
{Sanders}, D.~B., {Clemens}, D.~P., {Scoville}, N.~Z., \& {Solomon}, P.~M.
  1986, \apjs, 60, 1

\bibitem[{{Schuller} {et~al.}(2009){Schuller}, {Menten}, {Contreras},
  {Wyrowski}, {Schilke}, {Bronfman}, {Henning}, {Walmsley}, {Beuther},
  {Bontemps}, {Cesaroni}, {Deharveng}, {Garay}, {Herpin}, {Lefloch}, {Linz},
  {Mardones}, {Minier}, {Molinari}, {Motte}, {Nyman}, {Reveret}, {Risacher},
  {Russeil}, {Schneider}, {Testi}, {Troost}, {Vasyunina}, {Wienen}, {Zavagno},
  {Kovacs}, {Kreysa}, {Siringo}, \& {Wei{\ss}}}]{schuller2009}
{Schuller}, F., {Menten}, K.~M., {Contreras}, Y., {et~al.} 2009, \aap, 504, 415

\bibitem[{{Simon} {et~al.}(2001){Simon}, {Jackson}, {Clemens}, {Bania}, \&
  {Heyer}}]{simon2001}
{Simon}, R., {Jackson}, J.~M., {Clemens}, D.~P., {Bania}, T.~M., \& {Heyer},
  M.~H. 2001, \apj, 551, 747

\bibitem[{{Sridharan} {et~al.}(2002){Sridharan}, {Beuther}, {Schilke},
  {Menten}, \& {Wyrowski}}]{sridha}
{Sridharan}, T.~K., {Beuther}, H., {Schilke}, P., {Menten}, K.~M., \&
  {Wyrowski}, F. 2002, \apj, 566, 931

\bibitem[{{Stutzki} \& {Guesten}(1990)}]{stutzki1990}
{Stutzki}, J. \& {Guesten}, R. 1990, \apj, 356, 513

\bibitem[{{Thompson} {et~al.}(2006){Thompson}, {Hatchell}, {Walsh},
  {MacDonald}, \& {Millar}}]{thompson2006}
{Thompson}, M.~A., {Hatchell}, J., {Walsh}, A.~J., {MacDonald}, G.~H., \&
  {Millar}, T.~J. 2006, \aap, 453, 1003

\bibitem[{{Vassilev} {et~al.}(2008){Vassilev}, {Meledin}, {Lapkin}, {Belitsky},
  {Nystr{\"o}m}, {Henke}, {Pavolotsky}, {Monje}, {Risacher}, {Olberg},
  {Strandberg}, {Sundin}, {Fredrixon}, {Ferm}, {Desmaris}, {Dochev},
  {Pantaleev}, {Bergman}, \& {Olofsson}}]{vassilev2008}
{Vassilev}, V., {Meledin}, D., {Lapkin}, I., {et~al.} 2008, \aap, 490, 1157

\bibitem[{{Vasyunina} {et~al.}(2009){Vasyunina}, {Linz}, {Henning}, {Stecklum},
  {Klose}, \& {Nyman}}]{vasyunina2009}
{Vasyunina}, T., {Linz}, H., {Henning}, T., {et~al.} 2009, \aap, 499, 149

\bibitem[{{Walsh} {et~al.}(1998){Walsh}, {Burton}, {Hyland}, \&
  {Robinson}}]{walsh1998}
{Walsh}, A.~J., {Burton}, M.~G., {Hyland}, A.~R., \& {Robinson}, G. 1998,
  \mnras, 301, 640

\bibitem[{{Walsh} {et~al.}(2010){Walsh}, {Thorwirth}, {Beuther}, \&
  {Burton}}]{walsh2010}
{Walsh}, A.~J., {Thorwirth}, S., {Beuther}, H., \& {Burton}, M.~G. 2010,
  \mnras, 404, 1396

\bibitem[{{Williams} {et~al.}(1994){Williams}, {de Geus}, \&
  {Blitz}}]{williams1994}
{Williams}, J.~P., {de Geus}, E.~J., \& {Blitz}, L. 1994, \apj, 428, 693

\bibitem[{{Wilson}(1974)}]{wilson1974}
{Wilson}, T.~L. 1974, \aap, 31, 832

\bibitem[{{Wood} \& {Churchwell}(1989)}]{wc1989b}
{Wood}, D.~O.~S. \& {Churchwell}, E. 1989, \apjs, 69, 831

\bibitem[{{Woodward} {et~al.}(1984){Woodward}, {Helfer}, \&
  {Pipher}}]{woodward1984}
{Woodward}, C.~E., {Helfer}, H.~L., \& {Pipher}, J.~L. 1984, \mnras, 209, 209

\bibitem[{{Zhang} {et~al.}(2005){Zhang}, {Huang}, {Sun}, \& {Lu}}]{zhang2005c}
{Zhang}, Y., {Huang}, Y., {Sun}, J., \& {Lu}, D. 2005, \caa, 29, 9

\bibitem[{{Zinnecker} \& {Yorke}(2007)}]{zinnecker2007}
{Zinnecker}, H. \& {Yorke}, H.~W. 2007, \araa, 45, 481

\end{thebibliography}

\clearpage 

\begin{table*}[ht]
  \caption{Clump parameters for original data. The first part shows the coordinates, and the second and third parts the 875$\mu$m and the C$^{18}$O(2--1) data, respectively.}
\begin{tabular}{lrr|rrrrr|rrrrrr}
\hline \hline
\# & R.A. & Dec. & $S_{\rm{peak}}$ & $S_{\rm{int}}$ & $r_{\rm{eff}}$ & $N_{\rm{H}_2}$ & $M$ & $T_{\rm{peak}}$ & $\int T_{\rm{peak}}dv$ & $v_{\rm{peak}}$ & $\Delta v$ & $N_{\rm{H}_2}$ & $M_{\rm{vir}}$ \\
      &              &                   &        &       & & (dust)&       &     &     &      &     &(C$^{18}$O)& \\
& J2000.0 & J2000.0 & $\frac{\rm{Jy}}{\rm{beam}}$ & Jy & pc & $\frac{10^{22}}{\rm{cm}}$ & M$_{\odot}$ & K & K\,km\,s$^{-1}$ & km\,s$^{-1}$ & km\,s$^{-1}$ & $\frac{10^{22}}{\rm{cm}}$ & M$_{\odot}$ \\
\hline 
   1  &   18 08 55.83  &  -20 05 56.1  &  7.16  & 27.37 & 1.2  & 34.3  & 7359  & 2.8 & 7.2 & 11.2 & 2.4 & 2.8 & 285  \\ 
   2  &   18 08 59.96  &  -20 03 36.4  &  4.93  & 16.86 & 1.2  & 23.6  & 4534  & 1.3 & 2.6 & 10.1 & 1.8 & 1.0 & 167  \\ 
   3  &   18 09 01.61  &  -20 05 09.6  &  4.36  & 30.45 & 1.7  & 20.9  & 8187  & 2.1 & 6.5 & 11.9 & 2.9 & 2.5 & 416  \\ 
   4  &   18 09 24.34  &  -20 15 38.5  &  4.31  & 21.31 & 1.3  & 20.6  & 5731  &     &     &      &     &     &      \\ 
   5  &   18 08 49.22  &  -20 05 56.1  &  4.02  & 11.41 & 0.8  & 19.2  & 3067  & 0.8 & 4.2 & 11.7 & 4.8 & 1.6 & 1175 \\ 
   6  &   18 08 46.33  &  -20 05 50.2  &  3.81  & 18.61 & 1.3  & 18.2  & 5003  & 1.7 & 1.9 & 10.6 & 1.0 & 0.7 & 50   \\ 
   7  &   18 09 27.65  &  -20 19 08.1  &  3.68  & 27.65 & 1.3  & 17.6  & 7434  & 1.8 & 9.3 & 10.1 & 4.8 & 3.6 & 1159 \\ 
   8  &   18 09 21.45  &  -20 19 31.4  &  3.43  & 25.94 & 1.5  & 16.4  & 6974  & 2.5 & 11.7& 10.2 & 4.3 & 4.5 & 938  \\ 
   9  &   18 08 52.11  &  -20 06 07.8  &  2.37  & 10.44 & 1.1  & 11.3  & 2808  & 2.2 & 6.5 & 11.1 & 2.8 & 2.5 & 405  \\ 
  10  &   18 09 26.82  &  -20 17 29.1  &  2.33  & 15.49 & 1.3  & 11.1  & 4164  & 1.0 & 6.5 & 11.9 & 6.3 & 2.5 & 2014 \\ 
  11  &   18 09 25.58  &  -20 18 27.3  &  2.31  & 17.87 & 1.3  & 11.1  & 4804  & 1.0 & 7.6 & 12.9 & 6.9 & 2.9 & 2377 \\ 
  12  &   18 09 28.89  &  -20 16 48.3  &  2.30  & 18.61 & 1.4  & 11.0  & 5004  & 1.0 & 6.7 & 11.9 & 6.2 & 2.6 & 1909 \\ 
  13  &   18 09 23.09  &  -20 08 04.3  &  2.28  & 8.06  & 1.2  & 10.9  & 2168  &     &     &      &     &     &      \\ 
  14  &   18 09 03.26  &  -20 03 01.5  &  2.09  & 13.17 & 1.3  & 10.0  & 3540  & 0.8 & 1.2 & 11.2 & 1.4 & 0.5 & 101  \\ 
  15  &   18 09 00.36  &  -20 11 33.9  &  1.88  & 16.27 & 1.7  & 9.0   & 4374  &     &     &      &     &     &      \\ 
  16  &   18 09 20.62  &  -20 15 03.6  &  1.65  & 3.93  & 0.9  & 7.9   & 1055  &     &     &      &     &     &      \\ 
  17  &   18 09 35.92  &  -20 18 44.7  &  1.53  & 6.19  & 1.0  & 7.3   & 1665  & 0.8 & 2.6 &  9.7 & 3.1 & 1.0 & 469  \\ 
  18  &   18 09 33.85  &  -20 17 52.4  &  1.46  & 6.92  & 1.1  & 7.0   & 1860  &     &     &      &     &     &      \\ 
  19  &   18 08 47.56  &  -20 06 48.5  &  1.44  & 7.31  & 1.1  & 6.9   & 1967  & 1.0 & 3.4 &  9.7 & 3.1 & 1.3 & 499  \\ 
  20  &   18 09 35.56  &  -20 21 27.8  &  1.30  & 3.09  & 0.7  & 6.2   & 830   &     &     &      &     &     &      \\ 
  21  &   18 09 21.03  &  -20 16 07.6  &  1.26  & 11.90 & 1.4  & 6.0   & 3201  &     &     &      &     &     &      \\ 
  22  &   18 09 26.83  &  -20 21 22.0  &  1.22  & 4.29  & 0.9  & 5.8   & 1154  & 1.8 & 7.1 & 11.1 & 3.7 & 2.7 & 687  \\ 
  23  &   18 08 57.89  &  -20 07 06.0  &  1.08  & 5.35  & 1.1  & 5.2   & 1439  & 1.5 & 6.6 & 11.3 & 4.3 & 2.5 & 923  \\ 
  24  &   18 09 24.76  &  -20 20 47.1  &  1.03  & 5.53  & 1.0  & 4.9   & 1486  & 3.0 & 15.6& 11.4 & 4.8 & 6.0 & 1173 \\ 
  25  &   18 09 23.09  &  -20 09 31.6  &  1.03  & 4.39  & 0.9  & 4.9   & 1180  &     &     &      &     &     &      \\ 
  26  &   18 09 14.42  &  -20 18 56.5  &  0.99  & 5.32  & 1.1  & 4.7   & 1431  & 5.4 & 21.6& 12.7 & 3.7 & 8.3 & 702  \\ 
  27  &   18 08 41.36  &  -20 07 35.0  &  0.90  & 3.26  & 0.9  & 4.3   & 862   & 1.5 & 4.2 &  9.1 & 2.7 & 1.6 & 368  \\ 
  28  &   18 09 06.57  &  -20 03 18.9  &  0.89  & 4.22  & 1.1  & 4.3   & 1134  & 1.1 & 2.0 & 11.2 & 1.7 & 0.8 & 153  \\ 
  29  &   18 09 05.33  &  -20 03 42.2  &  0.89  & 9.02  & 1.4  & 4.3   & 2425  & 1.4 & 2.5 & 11.0 & 1.7 & 0.9 & 137  \\ 
  30  &   18 09 29.72  &  -20 20 06.3  &  0.87  & 2.82  & 0.8  & 4.2   & 758   & 0.9 & 8.2 & 10.3 & 8.1 & 3.1 & 3331 \\ 
  31  &   18 08 39.26  &  -20 18 56.3  &  0.83  & 2.06  & 0.8  & 4.0   & 553   &     &     &      &     &     &      \\ 
  32  &   18 09 21.02  &  -20 02 03.2  &  0.81  & 7.00  & 1.4  & 3.9   & 1881  &     &     &      &     &     &      \\ 
  33  &   18 09 39.23  &  -20 19 31.3  &  0.80  & 2.05  & 0.7  & 3.8   & 552   & 0.7 & 0.8 &  6.6 & 1.0 & 0.3 & 50   \\ 
  34  &   18 09 12.76  &  -20 17 46.6  &  0.77  & 2.09  & 0.8  & 3.7   & 563   & 1.3 & 11.0& 12.5 & 7.8 & 4.2 & 3099 \\ 
  35  &   18 09 34.69  &  -20 22 08.6  &  0.72  & 1.25  & 0.6  & 3.4   & 337   &     &     &      &     &     &      \\ 
  36  &   18 09 37.16  &  -20 16 42.4  &  0.71  & 4.90  & 1.2  & 3.4   & 1318  &     &     &      &     &     &      \\ 
  37  &   18 09 25.18  &  -20 25 44.1  &  0.70. & 4.88  & 1.3  & 3.3   & 1312  & 2.5 & 11.9& 11.2 & 4.6 & 4.6 & 1055 \\ 
  38  &   18 09 17.30  &  -20 07 11.9  &  0.69  & 2.80  & 0.9  & 3.3   & 752   & 0.9 & 2.6 & 13.4 & 2.6 & 1.0 & 347  \\ 
  39  &   18 09 35.51  &  -20 20 52.9  &  0.68  & 1.75  & 0.7  & 3.3   & 471   & 0.8 & 3.0 & 10.5 & 3.3 & 1.1 & 559  \\ 
  40  &   18 09 34.28  &  -20 22 26.0  &  0.67  & 1.02  & 0.5  & 3.2   & 273   & 0.5 & 1.4 &  9.8 & 2.6 & 0.6 & 331  \\ 
  41  &   18 08 59.95  &  -20 06 31.1  &  0.66  & 2.52  & 0.8  & 3.2   & 677   & 1.7 & 2.3 & 11.7 & 1.3 & 0.8 & 79   \\ 
  42  &   18 09 13.17  &  -20 18 04.1  &  0.66  & 1.85  & 0.7  & 3.2   & 498   & 1.7 & 10.4& 13.2 & 5.8 & 4.0 & 1722 \\ 
  43  &   18 09 31.80  &  -20 23 59.2  &  0.66  & 2.35  & 0.8  & 3.2   & 632   & 0.9 & 2.4 & 16.1 & 2.5 & 0.9 & 313  \\ 
  44  &   18 09 26.40  &  -20 14 28.6  &  0.65  & 3.57  & 1.0  & 3.1   & 961   & 0.5 & 1.3 &  9.8 & 2.2 & 0.5 & 255  \\ 
  45  &   18 09 37.99  &  -20 19 48.8  &  0.65  & 0.76  & 0.5  & 3.1   & 214   &     &     &      &     &     &      \\ 
  46  &   18 09 21.85  &  -20 10 00.7  &  0.64  & 1.72  & 0.7  & 3.1   & 461   &     &     &      &     &     &      \\ 
  47  &   18 09 20.21  &  -20 20 29.6  &  0.64  & 3.29  & 1.0  & 3.1   & 885   & 6.3 & 25.9& 12.2 & 3.9 & 10. & 747  \\ 
  48  &   18 08 55.40  &  -20 11 04.8  &  0.60  & 4.84  & 1.3  & 2.9   & 1301  & 1.2 & 2.3 & 14.0 & 1.8 & 0.9 & 158  \\ 
  49  &   18 08 36.36  &  -20 19 37.0  &  0.57  & 2.23  & 0.9  & 2.7   & 599   &     &     &      &     &     &      \\ 
  50  &   18 09 14.83  &  -20 15 32.7  &  0.56  & 1.97  & 0.8  & 2.7   & 529   &     &     &      &     &     &      \\ 
  51  &   18 09 33.02  &  -20 13 59.4  &  0.56  & 1.29  & 0.6  & 2.7   & 348   & 0.9 & 2.3 &  9.0 & 2.6 & 0.9 & 331  \\ 
  52  &   18 09 01.58  &  -20 30 23.6  &  0.56  & 3.43  & 1.1  & 2.7   & 923   & 0.9 & 1.4 & 11.7 & 1.4 & 0.5 & 100  \\ 
  53  &   18 08 45.08  &  -20 07 35.0  &  0.54  & 3.05  & 1.0  & 2.6   & 820   &     &     &      &     &     &      \\ 
  54  &   18 09 09.86  &  -20 29 37.1  &  0.53  & 1.43  & 0.7  & 2.5   & 385   & 0.8 & 2.9 & 10.7 & 3.4 & 1.1 & 568  \\ 
  55  &   18 08 41.36  &  -20 08 21.6  &  0.51  & 1.58  & 0.7  & 2.4   & 426   & 0.8 & 2.2 & 10.4 & 2.7 & 0.9 & 355  \\ 
  56  &   18 09 06.97  &  -20 26 36.5  &  0.50  & 1.76  & 0.8  & 2.4   & 474   & 0.8 & 0.9 & 11.7 & 1.1 & 0.4 & 66   \\ 
  57  &   18 08 31.40  &  -20 18 56.2  &  0.50  & 1.54  & 0.7  & 2.4   & 414   &     &     &      &     &     &      \\ 
  58  &   18 09 15.24  &  -20 16 36.7  &  0.48  & 1.65  & 0.7  & 2.3   & 444   & 1.0 & 4.0 &  9.7 & 3.9 & 1.5 & 762  \\ 
  59  &   18 09 11.52  &  -20 16 19.3  &  0.47  & 1.26  & 0.7  & 2.2   & 339   &     &     &      &     &     &      \\ 
  60  &   18 09 37.17  &  -20 20 23.7  &  0.47  & 0.57  & 0.4  & 2.2   & 154   &     &     &      &     &     &      \\ 
  61  &   18 08 40.54  &  -20 08 21.6  &  0.46  & 0.62  & 0.5  & 2.2   & 168   & 1.2 & 3.6 & 10.0 & 2.9 & 1.4 & 410  \\ 
  62  &   18 08 38.01  &  -20 20 58.6  &  0.45  & 0.88  & 0.6  & 2.2   & 231   &     &     &      &     &     &      \\ 
  63  &   18 09 09.04  &  -20 25 55.8  &  0.45  & 1.63  & 0.8  & 2.2   & 439   &     &     &      &     &     &      \\ 
  64  &   18 09 13.59  &  -20 16 19.3  &  0.45  & 0.88  & 0.6  & 2.2   & 238   &     &     &      &     &     &      \\ 
  65  &   18 09 41.28  &  -20 12 08.7  &  0.44  & 1.97  & 0.9  & 2.1   & 529   &     &     &      &     &     &      \\ 
  66  &   18 08 40.53  &  -20 08 33.2  &  0.43  & 1.08  & 0.6  & 2.1   & 289   &     &     &      &     &     &      \\ 
  67  &   18 08 45.91  &  -20 07 05.9  &  0.43  & 1.17  & 0.6  & 2.1   & 314   & 0.8 & 1.4 &  8.8 & 1.7 & 0.5 & 152  \\ 
  68  &   18 08 43.02  &  -20 06 54.3  &  0.43  & 2.49  & 0.9  & 2.1   & 669   & 1.1 & 3.6 & 15.8 & 3.2 & 1.4 & 505  \\ 
  69  &   18 09 15.66  &  -20 16 19.3  &  0.42  & 0.76  & 0.5  & 2.0   & 203   &     &     &      &     &     &      \\ 
  70  &   18 08 43.87  &  -19 58 56.8  &  0.41  & 0.93  & 0.6  & 2.0   & 249   &     &     &      &     &     &      \\ 
  71  &   18 09 06.55  &  -20 28 38.8  &  0.39  & 0.77  & 0.5  & 1.9   & 206   & 1.0 & 2.9 & 10.6 & 2.9 & 1.1 & 413  \\ 
  72  &   18 08 21.87  &  -20 22 54.8  &  0.32  & 0.45  & 0.4  & 1.5   & 121   &     &     &      &     &     &      \\ 
  73  &   18 09 26.42  &  -20 27 28.9  &  0.31  & 1.13  & 0.7  & 1.5   & 304   &     &     &      &     &     &      \\   
\hline \hline
\end{tabular}
\footnotesize{~\\ The table parameters are 875$\mu$m peak and integrated fluxes $S_{\rm{peak}}$ and $S_{\rm{int}}$, effective linear clump radius from the clumpfind search $r_{\rm{eff}}$, and calculated H$_2$ column densities $N_{\rm{H_2}}$ and gas masses $M$. The C$^{18}$O(2--1) part shows the peak and integrated line intensities $T_{\rm{peak}}$ and $\int T_{\rm{peak}}dv$, the peak velocities $v_{\rm{peak}}$ and line widths $\Delta v$, and the derived H$_2$ column densities $N_{\rm{H_2}}$ and virial gas masses $M_{\rm{vir}}$.}
\label{clump_parameters}
\end{table*}

\clearpage

\begin{table*}[ht]
  \caption{Clump parameters for $27.5''$ resolution. The first part shows the coordinates, and the second and third parts the 875$\mu$m and the C$^{18}$O(2--1) data, respectively.875\,$\mu$m clump parameters for $27.5''$ resolution.}
\begin{tabular}{lrr|rrrrrr|rrrrrr}
\hline \hline
\# & R.A. & Dec. & $S_{\rm{peak}}$ & $S_{\rm{int}}$ & $r_{\rm{eff}}$ & $N_{\rm{H}_2}$ & $M$ & $M_{\rm{peak}}$ & $T_{\rm{peak}}$ & $\int T_{\rm{peak}}dv$ & $v_{\rm{peak}}$ & $\Delta v$ & $N_{\rm{H}_2}$ & $M_{\rm{vir}}$ \\
      &              &              &      &     & & (dust)&    &     &     &      &      &     & (C$^{18}$O) & \\
& J2000.0 & J2000.0 & $\frac{\rm{Jy}}{\rm{beam}}$ & Jy & pc & $\frac{10^{22}}{\rm{cm}}$ & M$_{\odot}$ & M$_{\odot}$ & K & K\,km\,s$^{-1}$ & km\,s$^{-1}$ & km\,s$^{-1}$ & $\frac{10^{22}}{\rm{cm}}$ & M$_{\odot}$ \\
\hline 
   1  & 18 08 55.83 & -20 06 02.0 & 11.22& 36.16 & 1.7 & 26.2& 9723 & 3017 & 3.0 & 7.5  & 11.3 & 2.4 & 2.9 & 279  \\ 
   2  & 18 08 59.96 & -20 03 36.4 & 6.75 & 16.31 & 1.2 & 15.7& 4386 & 1815 & 1.5 & 2.6  & 10.0 & 1.7 & 0.9 & 141  \\ 
   3  & 18 09 24.34 & -20 15 38.5 & 6.71 & 31.44 & 2.0 & 15.6& 8454 & 1804 &     &      &      &     &     &      \\            
   4  & 18 09 01.61 & -20 05 09.6 & 6.67 & 34.26 & 2.2 & 15.6& 9213 & 1793 & 1.8 & 6.4  & 11.7 & 3.3 & 2.5 & 538  \\ 
   5  & 18 09 27.65 & -20 19 08.1 & 6.50 & 36.43 & 1.7 & 15.2& 9796 & 1749 & 1.8 & 9.0  & 10.0 & 4.7 & 3.5 & 1129 \\ 
   6  & 18 08 46.74 & -20 05 50.2 & 6.44 & 38.99 & 2.1 & 15.0& 1048 & 1732 & 1.1 & 2.6  & 10.5 & 2.2 & 1.0 & 244  \\ 
   7  & 18 09 21.45 & -20 19 31.4 & 5.68 & 33.06 & 2.1 & 13.3& 8890 & 1528 & 2.4 & 11.8 & 10.4 & 4.6 & 4.5 & 1052 \\ 
   8  & 18 09 26.41 & -20 17 40.8 & 4.27 & 23.00 & 1.7 & 10.0& 6183 & 1147 & 1.3 & 7.4  & 12.4 & 5.6 & 2.9 & 1568 \\ 
   9  & 18 09 28.89 & -20 16 48.3 & 3.77 & 21.92 & 1.8 & 8.8 & 5894 & 1015 & 1.5 & 8.1  & 12.0 & 5.1 & 3.1 & 1336 \\ 
  10  & 18 09 03.26 & -20 03 07.3 & 3.59 & 23.71 & 2.1 & 8.4 & 6376 & 965  & 0.6 & 0.9  & 11.3 & 1.4 & 0.3 & 99   \\ 
  11  & 18 09 00.36 & -20 11 33.9 & 3.22 & 16.32 & 1.9 & 7.5 & 4388 & 866  &     &      &      &     &     &      \\            
  12  & 18 09 23.09 & -20 08 04.3 & 3.10 & 7.36  & 1.3 & 7.2 & 1979 & 833  &     &      &      &     &     &      \\            
  13  & 18 09 36.34 & -20 18 44.7 & 2.44 & 6.03  & 1.1 & 5.7 & 1620 & 656  & 0.6 & 2.9  &  9.3 & 4.6 & 1.1 & 1074 \\ 
  14  & 18 09 33.85 & -20 17 52.4 & 2.34 & 11.23 & 1.8 & 5.5 & 3021 & 629  &     &      &      &     &     &      \\            
  15  & 18 09 20.62 & -20 15 03.6 & 2.17 & 11.43 & 1.8 & 5.1 & 3074 & 585  & 0.6 & 1.5  & 10.3 & 2.3 & 0.6 & 273  \\ 
  16  & 18 09 26.83 & -20 21 22.0 & 1.95 & 4.11  & 1.0 & 4.5 & 1105 & 524  & 2.1 & 7.7  & 11.2 & 3.5 & 3.0 & 602  \\ 
  17  & 18 09 35.52 & -20 21 27.8 & 1.89 & 7.37  & 1.5 & 4.4 & 1980 & 507  &     &      &      &     &     &      \\            
  18  & 18 08 58.30 & -20 07 06.0 & 1.81 & 7.10  & 1.4 & 4.2 & 1908 & 485  & 1.2 & 5.5  & 11.3 & 4.4 & 2.1 & 959  \\ 
  19  & 18 09 25.17 & -20 20 47.1 & 1.74 & 7.48  & 1.4 & 4.1 & 2011 & 469  & 3.2 & 16.2 & 11.5 & 4.8 & 6.2 & 1169 \\ 
  20  & 18 09 24.33 & -20 09 08.3 & 1.62 & 6.94  & 1.4 & 3.8 & 1865 & 436  &     &      &      &     &     &      \\                
  21  & 18 08 41.36 & -20 07 35.0 & 1.42 & 12.96 & 2.1 & 3.3 & 3483 & 381  & 1.2 & 3.6  &  9.2 & 2.8 & 1.4 & 386  \\ 
  22  & 18 09 21.02 & -20 02 03.2 & 1.33 & 7.63  & 1.7 & 3.1 & 2052 & 359  &     &      &      &     &     &      \\            
  23  & 18 09 39.23 & -20 19 31.3 & 1.15 & 3.38  & 1.1 & 2.7 & 910  & 309  & 0.6 & 0.9  &  6.7 & 1.5 & 3.5 & 108  \\ 
  24  & 18 09 13.17 & -20 17 52.4 & 1.09 & 10.21 & 2.0 & 2.5 & 2745 & 292  & 1.2 & 2.9  & 10.4 & 2.3 & 1.1 & 265  \\ 
  25  & 18 09 17.30 & -20 07 11.9 & 1.09 & 2.70  & 1.0 & 2.5 & 727  & 292  & 1.3 & 2.7  & 13.5 & 2.0 & 1.0 & 208  \\ 
  26  & 18 08 38.85 & -20 18 56.3 & 1.07 & 2.09  & 0.9 & 2.5 & 562  & 287  &     &      &      &     &     &      \\            
  27  & 18 09 32.21 & -20 23 59.2 & 1.03 & 2.88  & 1.1 & 2.4 & 775  & 276  & 0.9 & 2.7  & 15.9 & 2.8 & 1.1 & 385  \\ 
  28  & 18 09 25.18 & -20 25 44.1 & 1.01 & 4.63  & 1.5 & 2.3 & 1244 & 270  & 2.6 & 12.1 & 11.1 & 4.3 & 4.7 & 939  \\ 
  29  & 18 08 54.99 & -20 11 04.8 & 0.98 & 5.04  & 1.4 & 2.3 & 1356 & 265  & 2.6  & 14.3 & 2.3 & 1.0 & 262  \\ 
  30  & 18 09 01.58 & -20 30 29.4 & 0.94 & 3.57  & 1.2 & 2.2 & 960  & 253  &     &      &      &     &     &      \\            
  31  & 18 09 32.60 & -20 14 05.3 & 0.86 & 1.80  & 0.9 & 2.0 & 484  & 232  & 0.7 & 1.7  &  8.3 & 2.2 & 0.7 & 247  \\ 
  32  & 18 09 07.38 & -20 26 30.7 & 0.84 & 5.53  & 1.6 & 2.0 & 1486 & 226  &     &      &      &     &     &      \\   
  33  & 18 08 36.36 & -20 19 31.2 & 0.80 & 2.38  & 1.0 & 1.9 & 641  & 215  &     &      &      &     &     &      \\            
  34  & 18 08 32.23 & -20 19 13.7 & 0.78 & 2.01  & 1.0 & 1.8 & 540  & 210  &     &      &      &     &     &      \\            
  35  & 18 09 09.45 & -20 29 37.1 & 0.78 & 1.69  & 0.9 & 1.8 & 456  & 210  & 1.0 & 2.9  & 10.8 & 2.8 & 1.1 & 398  \\ 
  36  & 18 09 28.49 & -20 25 15.0 & 0.68 & 3.44  & 1.3 & 1.6 & 924  & 182  &     &      &      &     &     &      \\ 
  37  & 18 09 40.87 & -20 12 26.2 & 0.66 & 2.42  & 1.1 & 1.5 & 652  & 177  &     &      &      &     &     &      \\            
  38  & 18 08 38.01 & -20 20 58.6 & 0.64 & 1.34  & 0.9 & 1.5 & 361  & 171  &     &      &      &     &     &      \\            
  39  & 18 08 43.46 & -19 58 50.9 & 0.59 & 1.31  & 0.9 & 1.4 & 352  & 160  &     &      &      &     &     &      \\            
  40  & 18 08 21.87 & -20 22 54.8 & 0.51 & 0.60  & 0.6 & 1.2 & 160  & 138  &     &      &      &     &     &      \\            
  41  & 18 08 15.71 & -20 13 47.3 & 0.43 & 0.50  & 0.6 & 1.0 & 133  & 116  &     &      &      &     &     &      \\            
  42  & 18 08 03.82 & -20 01 33.3 & 0.41 & 0.34  & 0.5 & 1.0 & 91   & 110  &     &      &      &     &     &      \\            
  43  & 18 08 31.04 & -20 05 44.2 & 0.39 & 0.50  & 0.6 & 0.9 & 135  & 105  &     &      &      &     &     &      \\            
  44  & 18 09 38.35 & -20 00 53.2 & 0.39 & 0.35  & 0.5 & 0.9 & 93   & 105  &     &      &      &     &     &      \\             
\hline \hline
\end{tabular}
\footnotesize{The table parameters are 875$\mu$m peak and integrated fluxes $S_{\rm{peak}}$ and $S_{\rm{int}}$, effective linear clump radius from the clumpfind search $r_{\rm{eff}}$, and calculated H$_2$ column densities $N_{\rm{H_2}}$ and gas masses $M$ and M$_{\rm{peak}}$, derived from the total flux and peak flux, respectively. The C$^{18}$O(2--1) part shows the peak and integrated line intensities $T_{\rm{peak}}$ and $\int T_{\rm{peak}}dv$, the peak velocities $v_{\rm{peak}}$ and line widths $\Delta v$, and the derived H$_2$ column densities $N_{\rm{H_2}}$ and virial gas masses $M_{\rm{vir}}$.}
\label{clump_parameters_smooth}
\end{table*}

\end{document}